\documentclass[pra,twocolumn,
superscriptaddress,
showpacs,
aps
]{revtex4-1}
\usepackage{graphicx}
\usepackage{amsmath}
\usepackage{amsfonts}
\usepackage{amssymb}
\usepackage{hyperref}
\usepackage{color}

\newcommand{\ket}[1]{|{#1}\rangle}

\newcommand{\beq}{\begin{equation}}
\newcommand{\eeq}{\end{equation}}

\usepackage{dsfont}
\usepackage{placeins} 
\usepackage{bm}           

\begin{document}

\title{Superradiant optomechanical phases of cold atomic gases in optical resonators}
\author{Simon B. J\"ager} 
\address{Theoretische Physik, Universit\"at des Saarlandes, D-66123 Saarbr\"ucken, Germany} 
\author{Murray J. Holland} 
\address{JILA, National Institute of Standards and Technology and Department of Physics,
	University of Colorado, Boulder, Colorado 80309-0440, USA} 
\address{Center for Theory of Quantum Matter, University of Colorado, Boulder, Colorado 80309, USA} 
\author{Giovanna Morigi} 
\address{Theoretische Physik, Universit\"at des Saarlandes, D-66123 Saarbr\"ucken, Germany}

\date{\today}

\begin{abstract}
We theoretically analyze superradiant emission of light from a cold atomic gas, when mechanical effects of photon-atom interactions are considered. The atoms are confined within a standing-wave resonator and an atomic metastable dipolar transition couples to a cavity mode. The atomic dipole is incoherently pumped in the parameter regime that would correspond to stationary superradiance in absence of inhomogeneous broadening. Starting from the master equation for cavity field and atomic degrees of freedom we derive a mean-field model that allows us to determine a threshold temperature, above which thermal fluctuations suppress superradiant emission. We then analyze the dynamics of superradiant emission when the motion is described by a mean-field model. In the semiclassical regime and below the threshold temperature we observe that the emitted light can be either coherent or chaotic, depending on the incoherent pump rate. We then analyze superradiant emission from an ideal Bose gas at zero temperature when the superradiant decay rate $\Lambda$ is of the order of the recoil frequency $\omega_R$. We show that the quantized exchange of mechanical energy between the atoms and the field gives rise to a  threshold, $\Lambda_c$, below which superradiant emission is damped down to zero. When $\Lambda>\Lambda_c$ superradiant emission is accompanied by the formation of matter-wave gratings diffracting the emitted photons. The stability of these gratings depends on the incoherent pump rate $w$ with respect to a second threshold value $w_c$. For $w>w_c$ the gratings are stable and the system achieves stationary superradiance. Below this second threshold the coupled dynamics becomes chaotic. We characterize the dynamics across these two thresholds and show that the three phases we predict (incoherent, coherent, chaotic) can be revealed via the coherence properties of the light at the cavity output. 
\end{abstract}

\maketitle
	
\section{Introduction}
Superradiance is a quantum interference phenomenon in the emission amplitudes of an ensemble of dipoles \cite{Dicke:1954,Gross:1982}, which is accompanied by a macroscopic coherence within the ensemble \cite{Dicke:1954}. Superradiant enhancement is intimately related to spontaneous synchronization of quantum systems \cite{Zambrini:2016} and can also be observed when a collection of dipoles is spatially confined within their resonance wave length and/or when they interact with the resonant mode of a cavity \cite{Kim:2018,Angerer:2019,Laske:2019}. In these settings lasing at the frequency and linewidth of the collective dipole can be observed: this regime has been also denoted by stationary superradiance \cite{Meiser:2009,Meiser:2010:1,Meiser:2010:2,Bohnet:2012,Tieri:preprint,Debnath:2018} and can reach ultra narrow linewidths \cite{Meiser:2009,Bohnet:2012,Maier:2014,Tieri:preprint,Norcia:2016:1,Norcia:2016:2,Debnath:2018}. The dynamical properties can be cast in terms of a Kuramoto model \cite{Mori:1998,Acebron:2005,Gupta:2018,Zhu:2015,Tucker:2018}. When the collective oscillations are sustained by the interplay between an incoherent pump and the long-range interaction between the dipoles, the collective dipole locks to a frequency determined by the rate of the incoherent pump. In this case the dynamics exhibits the principal features of a time crystal \cite{Tucker:2018} in a driven-dissipative setup \cite{Iemini:2018,Gong:2018,Kessler:2019}. Recent works have analyzed stationary superradiance in solid-state, inhomogeneously broadened environments \cite{Debnath:2019}. 

\begin{figure}[h!]
	\center \includegraphics[width=0.85\linewidth]{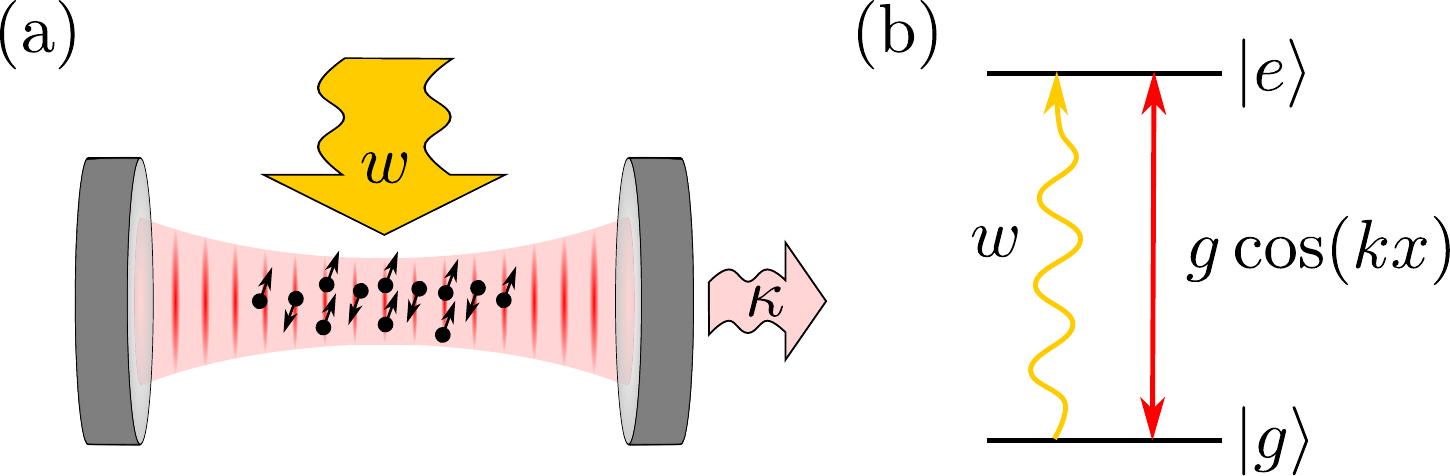}
	\caption{(a) A cold atomic gas is confined within a standing-wave resonator that decays at rate $\kappa$. Photons are emitted into the cavity by a metastable dipolar transition, which is broadened by the cavity field and which is incoherently pumped at rate $w$. We analyze the properties of the light when the mechanical effect of cavity photon-atom interactions is considered.  (b) Sketch of the relevant atomic levels and their coupling rates. The ground and excited states are $|g\rangle$ and $|e\rangle$, respectively, $g\cos(kx)$ is the strength of the coupling with the resonator, which is modulated by the spatial mode with wavenumber $k$ along $x$. The incoherent pump $w$ is effectively realized by optical pumping via a third level, as experimentally demonstrated in Ref.~\cite{Bohnet:2012}. \label{Fig:1}}
\end{figure}

\begin{figure}[h!]
	\center \includegraphics[width=0.9\linewidth]{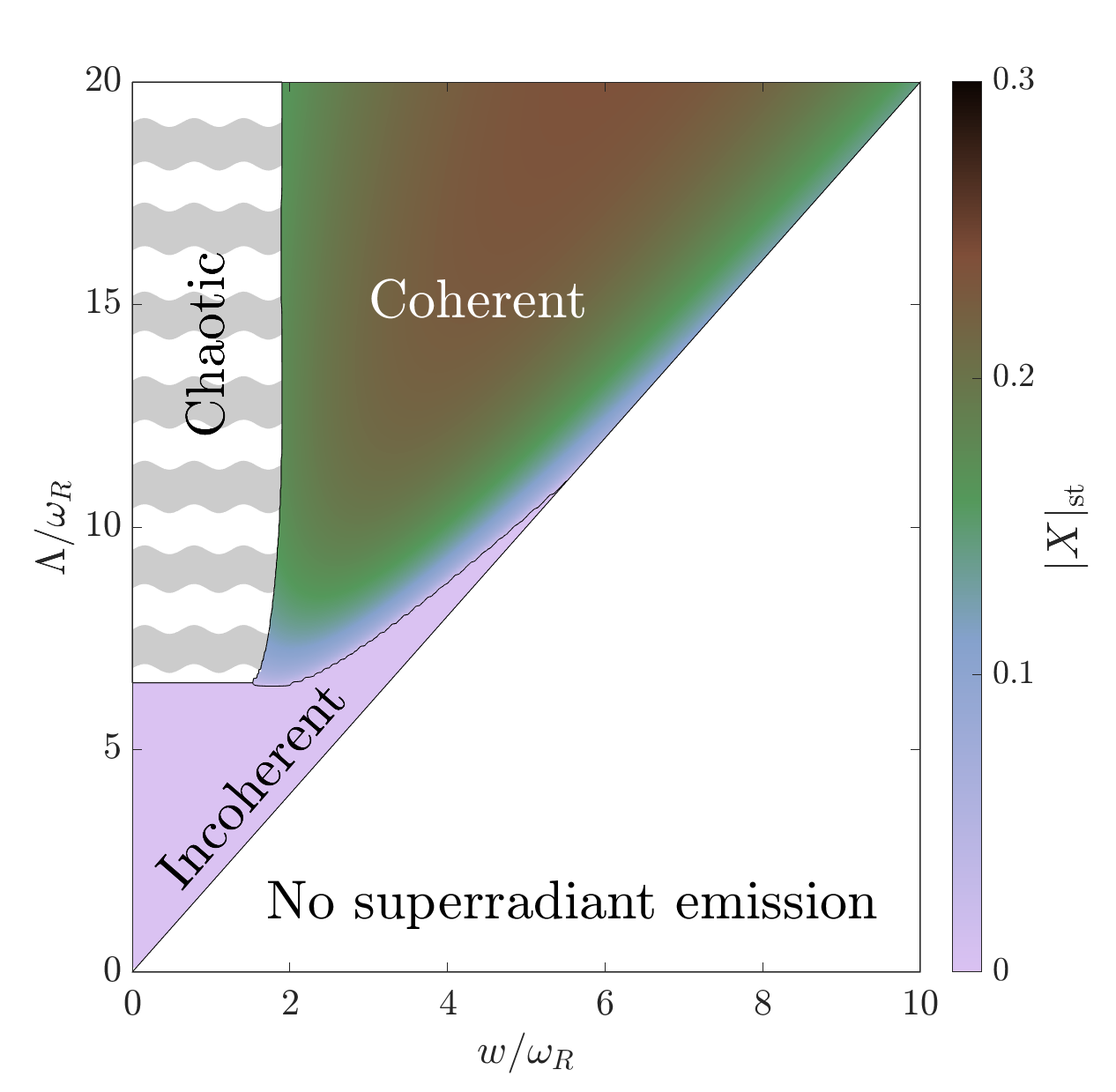}\\
	\caption{Phases due to the optomechanical dynamics in the $w-\Lambda$ plane (rates in units of $\omega_R$). The phases are labeled by the corresponding coherence properties of the emitted light, the color scale in the coherent and incoherent phases indicates the value of the order parameter $X$ (and thus of the emitted field) at steady state. The initial state is an ideal Bose gas at temperature $T=0$, and the phases are found for the asymptotic state of the dynamics described by Eq.~\eqref{MEq:MF} for $\Delta=\kappa/2$. In the absence of optomechanical coupling one also finds the same region of ``no superradiant emission'', while the other half of the diagram would be characterized by coherent light whose properties are independent of the ratio $\Lambda/\omega_R$ and $w/\omega_R$. See Sec.~\ref{Sec:BEC}.
\label{phasediagram}}
\end{figure}

In this work we analyze the stability and dynamics of stationary superradiance when the emitters are atoms or molecules whose dipolar transitions couple to the mode of a lossy standing-wave resonator. The setup we consider is illustrated in Fig.~\ref{Fig:1}. Here, the atoms are incoherently pumped, and therefore no coherence is established by the process pumping energy into the system. The system parameters are in the regime where stationary superradiant emission (SSR) is predicted \cite{Meiser:2009}. In these settings, theoretical studies based on semiclassical models predicted cooling of the atomic motion to ultra-low temperatures \cite{Xu:2016,Jaeger:2017,Hotter:2019}, and in particular to the regime in which the atoms form correlations between internal and external degrees of freedom that allow them to synchronize \cite{Jaeger:2017}. In the present work we derive and discuss a model which allows us to study the effect of semiclassical and quantum fluctuations on stationary superradiance. The present study is based and complements the work presented in Ref.~\cite{Jaeger:2019}. We analyze in detail the regime of validity of the quantum mean-field model of Ref.~\cite{Jaeger:2019}. In the regime of validity we then analyze in detail the phases emerging from the optomechanical coupling with collective spin and atomic motion, which are summarized in the diagram of Fig.~\ref{phasediagram}. This diagram shows the asymptotic state found when atoms forming a Bose-Einstein condensate are coupled to a resonator in the setup of Fig.~\ref{Fig:1}. The axes of the diagram are the incoherent pump rate $w$ and the superradiant decay rate $\Lambda$. The standard regime of SSR is for $w<\Lambda$, which in the diagram is above the diagonal line, while the single atom linewidth is here assumed to be much smaller than the recoil frequency $\omega_R$. The optomechanical coupling reduces the parameter region where SSR occurs, and gives rise to novel, chaotic-like behavior.  In the present article we analyze in detail the atoms and light properties across these phases.

This manuscript has the following structure. In Section~\ref{sec:2} we present the derivation of the quantum mean-field model starting from the full master equation and determine the regime of validity of the mean-field treatment. In Sec. \ref{sec:3} we determine the steady state by means of a stability analysis. Drawing from this model we determine the stationary state and the dynamics that leads to it for a thermal gas. In Sec.~\ref{Sec:BEC} we assume that the atoms initially form a Bose-Einstein condensate and study the dynamics across the coherent-incoherent and coherent-chaotic phases of the phase diagram in Fig.~\ref{phasediagram}. We then discuss the experimental parameter regimes where these dynamics can be observed. The conclusions are drawn in Sec.~\ref{Sec:conclusions}, while the appendices provide details of the calculations presented in Secs.~\ref{sec:2}, \ref{sec:3}, and \ref{Sec:BEC}.

\section{Derivation of a quantum mean field model}\label{sec:2}

In this section we introduce a model consisting of $N$ atoms whose dipolar transition couples to a single-mode cavity and is incoherently driven by an external pump field. The scheme, as sketched in Fig.~\ref{Fig:1}, contains dynamics that includes the mechanical effects of resonant light in the cavity standing-wave field on the atomic motion. We then derive an effective mean-field model that makes up the basis of our studies.

\subsection{Master equation}

We consider a gas of $N$ atoms or molecules of mass $m$ confined within an optical resonator. The relevant internal atomic degrees of freedom are the ground state $|g\rangle$ and the metastable excited state $|e\rangle$, which form a dipolar transition at frequency $\omega_a$. The dipole is transversally pumped by an incoherent field at rate $w$ and couples to the standing-wave mode of a cavity at frequency $\omega_c$ and wave vector $\vec k=k \vec e_x$. Here, $\vec e_x$ is the unit vector pointing along the $x$ (cavity) axis. In the following we assume that the atomic motion is considered to be tightly confined along the cavity axis, so that we restrict our analysis to one dimension along $\vec e_x$. We denote the canonically conjugate positions and momenta of the atoms by $\hat x_j$ and $\hat p_j$, respectively ($j=1,\ldots, N$), with the usual commutation relation $[\hat{x}_l,\hat{p}_m]={\rm i}\hbar \delta_{l,m}$. 

The dynamics of the density operator $\hat{\rho}$ of the cavity mode and of the atoms' internal and external degrees of freedom results from the interplay between the coherent interactions and the incoherent processes. This is governed by the Born-Markov master equation
\begin{align}
\frac{\partial \hat{\rho}}{\partial t}=\frac{1}{{\rm i}\hbar}[\hat{H},\hat{\rho}]+\kappa\mathcal{L}[\hat{a}]\hat{\rho}+w\sum_{j=1}^N\mathcal{L}[\hat{\sigma}_j^{\dag}]\hat {\rho}\,, \label{Mastereq0}
\end{align}
where the first term on the right-hand side describes the coherent dynamics, while the second and third terms describe cavity dissipation and incoherent atom pumping, respectively. Here, $\kappa$ is the cavity loss rate, $w$ is the rate of incoherent pumping, and the dissipative superoperator is given in usual form by
\begin{align*}
\mathcal{L}[\hat{O}=\hat a, \hat{\sigma}^{\dag}_j]\hat\rho=-\frac{1}{2}\left(\hat{O}^{\dag}\hat{O}\hat{\rho}+\hat{\rho}\hat{O}^{\dag}\hat{O}-2\hat{O}\hat{\rho}\hat{O}^{\dag}\right)\,.
\end{align*}
The field operators $\hat{a}$ and $\hat{a}^{\dag}$ annihilate and create, respectively, a cavity photon, with $[\hat{a},\hat{a}^\dag]=1$. The atomic operators $\hat{\sigma}_j=|g\rangle_j\langle e|$ and $\hat{\sigma}_j^\dag=|e\rangle_j\langle g|$ de-excite and excite, respectively, the electron of the $j$th atom. 

In the absence of incoherent processes, the total energy of atoms and cavity photons is composed of the sum of the atoms, field, and interaction energy (in the electric-dipole and rotating-wave approximation), and is described by the Hamiltonian 
\begin{align}
\hat{H}=&\sum_{j=1}^N\frac{\hat{p}_j^2}{2m}+\hbar\Delta\hat{a}^{\dag}\hat{a}+\hbar \frac{g}{2}\sum_{j=1}^N\cos(k\hat{x}_j)(\hat{a}^{\dag}\hat{\sigma}_{j}+\mathrm{H.c.})\,,\label{H0}
\end{align}
which is here reported in the interaction picture rotating at the atomic frequency $\omega_a$. In this reference frame the free atomic energy is the kinetic energy, while the cavity frequency is shifted by $\omega_a$, and is given by the cavity-atom detuning $\Delta=\omega_c-\omega_a$. The coupling with photons and atoms is scaled by the vacuum Rabi frequency $g$ and is sinusoidally modulated by the spatial envelope of the cavity field at position $x_j$ according to the mode-function $\cos(k x)$. This modulation can be included in the definition of the collective dipole $\hat{X}$ that couples to the cavity field:
\begin{align}
\hat{X}=&\frac{1}{N}\sum_{j=1}^N\cos(k\hat{x}_j)\hat{\sigma}_j\,.\label{X0}
\end{align} 
Because of thermal and/or quantum fluctuations the amplitudes $\cos(k\hat{x}_j)$ are dynamical variables, therefore the decomposition of the collective dipole into single atomic excitations varies with time. 

\subsection{Time-scale separation and effective dynamics}

In the regime in which the cavity dissipation rate is the largest frequency characterizing the dynamics, it is possible to adiabatically eliminate the cavity degrees of freedom from the dynamics of the atoms. This results in a long-range atom-atom force that has a dispersive and a dissipative component \cite{Larson:2008,Schuetz:2013}. For the case in which the atoms' internal degrees of freedom are described by spins, the resulting master equation reads \cite{Jaeger:2019}:
\begin{align}
\frac{\partial\hat{\rho}_{N}}{\partial t}=\frac{1}{{\rm i}\hbar}[\hat{H}_{\mathrm{eff}},\hat{\rho}_{N}]+N\Lambda\mathcal{L}[\hat{X}]\hat{\rho}_{N}+w\sum_{j=1}^N\mathcal{L}[\hat{\sigma}_j^{\dag}]\hat {\rho}_{N}\,. \label{Mastereq1}
\end{align}
Herein the coherent dynamics is governed by the effective Hamiltonian
\begin{align}
\hat{H}_{\mathrm{eff}}=&\sum_{j=1}^N\frac{\hat{p}_j^2}{2m}-\frac{\hbar N\Lambda}{2}\frac{\Delta}{\kappa/2}\hat{X}^{\dag}\hat{X},\label{H1}
\end{align}
and we have introduced the characteristic frequency parameter
\begin{align}
\Lambda=&\frac{Ng^2\kappa/4}{\Delta^2 +\kappa^2/4}\,.\label{Lambda}
\end{align}
Equation~\eqref{H1} illustrates that in this regime the total energy is a sum of the kinetic energies of every atom and a collective spin-spin coupling that is described by $\hat{X}^{\dag}\hat{X}$. The incoherent dynamics shown in Eq.~\eqref{Mastereq1} is governed by both the incoherent repumping of every atom and by the collective decay of $\hat{X}$ with rate $\Lambda$. 
In Appendix~\ref{App:A} we report the details of the derivation starting from Eq.~\eqref{Mastereq1} using the projector method  \cite{Gardiner:QuantumNoise}. 

We here remark that the theoretical perturbation treatment, which leads to Eq.~\eqref{Mastereq1}, is based on the assumption that the cavity field dynamics follows adiabatically the state of the atom. In this limit, by solving the Heisenberg-Langevin equations corresponding to the dynamics of Eq.~\eqref{Mastereq0} in the Schr\"odinger picture, the field operator $\hat a$ takes the form~\cite{Jaeger:2017}
\begin{equation}
\hat a(t)\approx-{\rm i}\frac{gN/2}{{\rm i}\Delta+\kappa/2}\hat X(t)+\hat{\mathcal{F}}(t)\,.
\end{equation}
Here $\hat{\mathcal F}(t)$ describes the quantum noise within a coarse-grained time-interval treatment, such that $\langle \hat{\mathcal{F}}\rangle=0$ and only two-time correlators can be non-zero \cite{Habibian:2013,Jaeger:2017}. Hence, for $\langle \hat X(t)\rangle \neq 0$ the electric field amplitude does not vanish, and $\langle \hat a(t)\rangle\propto \langle \hat X(t)\rangle\neq 0$. For $\langle \hat X(t)\rangle = 0$, instead, the field mean value vanishes and its intensity is dominated by shot-noise fluctuations. In this case the light that is produced is incoherent. 

\subsection{Mean-field model}\label{sec:6}

Even though, as we have just shown, it is possible to eliminate the cavity degrees of freedom, the full dynamics described by Eq.~\eqref{Mastereq1} is still intractable. The global-range interactions, on the other hand, motivate application of a mean-field approach. In order to justify this approach we first analyze the BBGKY (Bogoliubov-Born-Green-Kirkwood-Yvon) hierarchy, truncating at the second order level of the correlation hierarchy. For this purpose, starting from the master equation for the $N$-particle density matrix, Eq.~\eqref{Mastereq1}, we first determine the master equation for the $\ell$-particle density matrix $ \hat{\rho}_\ell$ with $1\le \ell\le N$. This density matrix is defined in the $\ell$-particle Hilbert space $\mathcal{H}^{\otimes \ell}$ and is obtained by tracing out the degrees of freedom of the $(N-\ell)$-particle Hilbert spaces $\mathrm{Tr}_{N-\ell}(\hat{\rho}_N)=\hat{\rho}_{\ell}$. Its master equation is derived by applying $\mathrm{Tr}_{N-\ell}$ onto both sides of Eq.~\eqref{Mastereq1}, to give:
\begin{align}
\frac{\partial \hat{\rho}_\ell}{\partial t}=\frac{1}{{\rm i}\hbar}[\hat{H}_\ell,\hat{\rho}_\ell]+\frac{\Lambda}{N}{\mathcal L}[\hat J_\ell]\hat\rho_\ell+\sum_{j=1}^\ell w{\mathcal L}[\hat \sigma_j]\hat\rho_\ell
+\mathcal{D}_\ell[\hat{\rho}_{\ell+1}]\label{Masterequl}
\end{align}
where the $\ell$-particle Hamiltonian reads:
\begin{align}
\hat{H}_{\ell}=&\sum_{j=1}^\ell\frac{\hat{p}_j^2}{2m}-\frac{\hbar \Delta}{2(\kappa/2)}\frac{\Lambda}{N}\hat{J}_\ell^{\dag}\hat{J}_\ell\,,\label{couplingrhol1}
\end{align}
with 
\begin{align}
\hat{J}_\ell=\sum_{j=1}^\ell\hat{\sigma}_j\cos(k\hat{x}_j).\label{Jl}
\end{align}
Moreover, $\hat\rho_\ell$ is coupled to $\hat\rho_{\ell+1}$ by means of the superoperator:
\begin{align}
\mathcal{D}_\ell[\hat{\rho}_{\ell+1}]=&{\rm i}\left(1-\frac{l}{N}\right)\frac{\Lambda}{2\sin\chi}{\rm e}^{-{\rm i}\chi}[\hat{J}_\ell,\hat{\mathcal{X}}_\ell^*[\hat{\rho}_{\ell+1}]] +\mathrm{H.c.}\,,\label{couplingl+1}
\end{align}
with 
\begin{align}
\tan\chi=\frac{\kappa}{2\Delta}\label{tanchi}
\end{align}
 and 
\begin{align}
\hat{\mathcal{X}}_\ell[\hat{\rho}_{\ell+1}]=\mathrm{Tr}^{(\ell+1)}\left(\hat{\sigma}_{\ell+1}\cos(k\hat{x}_{\ell+1})\hat{\rho}_{\ell+1}\right)
\end{align}
The linear mapping $\mathrm{Tr}^{(k)}$ denotes the trace over the $k$th degree of freedom.

We now consider the equation for $\ell=1$. The single-particle density matrix $\hat{\rho}_1$ couples to the two-particle density matrix $\hat{\rho}_2$. The latter can be decomposed as
\begin{align}
\hat{\rho}_2=\hat{\rho}_1\otimes\hat{\rho}_1+\hat{g}_2,\label{decompositionrho2}
\end{align}
where the first term is a mean-field term and $\hat{g}_2$ describes correlations beyond mean field. The master equation for $\hat \rho_1$ then can be cast into the form:
\begin{align}
\frac{\partial \hat{\rho}_1}{\partial t}=&\mathcal{L}_{\mathrm{mf}}[\hat{\rho}_1]\hat{\rho}_1+\mathcal{L}_2[\hat{g}_2],\label{rho1dynamics}
\end{align}
where the term first term on the right-hand side gives the mean-field master equation, while the second term gives the beyond-mean-field corrections. In detail,
\begin{align}
\mathcal{L}_{\mathrm{mf}}[\hat{\rho}_1]\hat{\rho}_1=&\frac{1}{{\rm i}\hbar}[\hat{H}_{\mathrm{mf}},\hat{\rho}_{1}]+\frac{\Lambda}{N}\mathcal{L}[\hat{\sigma}\cos(k\hat{x})]\hat{\rho}_{1}+w\mathcal{L}[\hat{\sigma}^{\dag}]\hat {\rho}_{1}, \label{Masterequation3}
\end{align}
with the effective Hamiltonian 
\begin{align}
\hat{H}_{\mathrm{mf}}=&\frac{\hat{p}^2}{2m}-\frac{\hbar \Lambda}{2N\tan\chi}\hat{\sigma}^{\dag}\hat{\sigma}\cos^2(k\hat{x})\\
&-\frac{\hbar\Lambda}{2\sin\chi}\left({\rm e}^{-{\rm i}\chi} X[\hat{\rho}_1]^*\hat{\sigma}+{\rm e}^{{\rm i}\chi} X[\hat{\rho}_1]\hat{\sigma}^{\dag}\right)\cos(k\hat{x})\,.\label{H2}
\end{align}
This mean-field Hamiltonian is composed of a non-linearity that depends on the state of the system through the mean-field order parameter
\begin{align}
X[\hat{\rho}_1]=\mathrm{Tr}\{\hat{\sigma}\cos(k\hat{x})\hat{\rho}\}\,.\label{mforderparameter}
\end{align}
The equation of motion for the beyond-mean-field corrections $\hat g_2$ is reported in Appendix~\ref{App:B}.

The dynamics of the system can be described by a mean-field theory when the coupling to $\hat g_2$ in Eq.~\eqref{rho1dynamics} can be discarded. For this purpose we analyze now the equation of motion of $\hat{g}_2$ and make some considerations on its order of magnitude, assuming that $\hat{g}_2(0)=0$ at $t=0$, or more explicitly that  $\hat{\rho}_N(0)=\hat{\rho}_1(0)\otimes \ldots \otimes\hat{\rho}_1(0)$. For this initial state (see Appendix~\ref{App:B})
\begin{align}
\frac{\partial \hat{\rho}_1}{\partial t}=&\mathcal{L}_{\mathrm{mf}}[\hat{\rho}_1]\hat{\rho}_1+\mathcal{L}_1[\hat{g}_{2}],\label{rho1eq}\\
\frac{\partial \hat{g}_2}{\partial t}=&(\mathrm{id}\otimes\mathcal{L}_{\mathrm{mf}}[\hat{\rho}_1]+\mathcal{L}_{\mathrm{mf}}[\hat{\rho}_1]\otimes \mathrm{id})\hat{g}_2\nonumber\\
&+{\rm i}\frac{\Lambda}{2\sin\chi}\left({\rm e}^{-{\rm i}\chi}\hat{\mathcal{X}}_1^*[\hat{g}_2]\otimes [\hat{J}_1,\hat{\rho}_1] -\mathrm{H.c.}\right)\nonumber\\
&+{\rm i}\frac{\Lambda}{2\sin\chi}\left({\rm e}^{-{\rm i}\chi}[\hat{J}_1,\hat{\rho}_1]\otimes\hat{\mathcal{X}}_1^*[\hat{g}_2] -\mathrm{H.c.}\right)\nonumber\\
&+{\rm i}\frac{\Lambda}{2N\sin\chi}\left({\rm e}^{-{\rm i}\chi}[\hat{J}_1,\hat{\rho}_1]\otimes \hat{\rho}_1(\hat{J}_1^{\dag}-X^*\hat{1})-\mathrm{H.c.}\right)\nonumber\\
&+{\rm i}\frac{\Lambda}{2N\sin\chi}\left({\rm e}^{-{\rm i}\chi} \hat{\rho}_1(\hat{J}_1^{\dag}-X^*\hat{1})\otimes[\hat{J}_1,\hat{\rho}_1]-\mathrm{H.c.}\right)\nonumber\\
&+{\rm i}\frac{\Lambda}{2\sin\chi}\left({\rm e}^{-{\rm i}\chi}[\hat{J}_2,\hat{\mathcal{X}}_2^*[\hat{g}_3]] -\mathrm{H.c.}\right)\,,\label{g2eq}
\end{align}
where $\mathrm{id}$ is the identity superoperator that maps every operator on itself, $\hat{1}$ is the unity operator and $X=X[\hat{\rho}_1]$. The order of magnitude of the individual terms on the right-hand side suggests that, under the assumption that $\hat{g}_2(0)=0$ then $\hat{g}_2(t)\sim {\rm O}(1/N)$ for times $t\ll N/\Lambda$. Since the characteristic timescale for the evolution of $\hat{\rho}_1$ is $\Lambda^{-1}$, we can then identify a timescale separation for $N\to \infty$ and assume that $\hat{g}_2$ remains zero over the time-scale on which we analyze the mean-field dynamics. This requires us to choose the thermodynamic limit where $\Lambda$ is constant as $N\to\infty$. These considerations and analysis based on the mean-field model are founded on this assumption and thus are only valid for timescales smaller than $N/\Lambda$.

\section{Mean-field analysis}\label{sec:3}

We now consider the regime of validity of the mean-field model and analyze the predictions of the mean-field master equation for the single-particle density matrix $\hat \rho_1$:
\begin{align}
\frac{\partial \hat{\rho}_1}{\partial t}=&\frac{1}{{\rm i}\hbar}[\hat{H}_{\mathrm{mf}},\hat{\rho}_{1}]+\frac{\Lambda}{N}\mathcal{L}[\hat{\sigma}\cos(k\hat{x})]\hat{\rho}_{1}+w\mathcal{L}[\hat{\sigma}^{\dag}]\hat {\rho}_{1}\,, \label{MEq:MF}
\end{align}
where $\hat{H}_{\mathrm{mf}}$ is given in Eq.~\eqref{H2}. We first analyze the stationary states, namely, the solutions of Equation
$$\partial_t\hat{\rho}_1=0\,,$$
keeping in mind that this is strictly valid for $N\to \infty$. For finite, but large $N$, the single-particle density matrix asymptotically approaches the quasi-stationary solution, which is consistent with our approximation, as long as the timescale on which they reach this is smaller than $N/\Lambda$. We then perform the stability analysis of the states we identify and determine a phase diagram.

\subsection{Mean-field energy}

We make first some preliminary considerations by studying the dynamics of the mean-field energy of the system. The mean value is given by
\begin{align}
\label{Hmf:mean}
\langle \hat{H}_{\mathrm{mf}}\rangle=\frac{\langle\hat{p}^2\rangle}{2m}-\hbar\frac{\Lambda}{\tan\chi}|X(t)|^2\,,
\end{align}
where $\langle \hat O\rangle={\rm Tr}\{\hat O\hat\rho_1\}$ and the order parameter depends on the quantum state, and thus on time. Its time evolution is given by the equation
\begin{align}
\frac{d\langle \hat{H}_{\mathrm{mf}}\rangle}{dt}
=&\mathrm{Tr}\left\{\hat{H}_{\mathrm{mf}}\frac{\partial\hat{\rho}}{\partial t}\right\}+\mathrm{Tr}\left\{\frac{\partial \hat{H}_{\mathrm{mf}}}{\partial t}\hat{\rho}\right\}\,
\label{energychange}\,.
\end{align}
We then substitute Eq.~\eqref{Hmf:mean} into the left-hand side and use Eq.~\eqref{MEq:MF} to expand the right-hand side. By reordering the resulting terms we can cast Eq.~\eqref{energychange} into an equation for the dynamics of the mean kinetic energy, which reads
\begin{align}
\frac{d}{dt}\frac{\langle \hat{p}^2\rangle}{2m}
=&\frac{\hbar\Lambda}{2}\left[\left(\frac{w}{\tan\chi}+2\frac{d\phi}{dt}\right)|X|^2+\frac{\Delta}{\kappa/2}\frac{d|X|^2}{dt}\right], \label{kineticdyn}
\end{align}
where we have used the notation $X(t)=|X(t)|e^{{\rm i}\phi(t)}$. We thus note that the dynamics of the mean kinetic energy is determined both by the time evolution of both the amplitude and phase of the order parameter. 

\subsection{Stationary states}

The nonlinear dependence of the coefficients of the Hamiltonian in Eq.~\eqref{H2} on $\hat{\rho}_1$ requires a careful approach in determining the stationary solution (assuming that this exists). We choose to first look for a class of solutions for which $X[\hat{\rho}_1]=0$, namely, where the non-linearity vanishes. This class of solutions corresponds to stationary states for which the expectation value of the cavity field vanishes and the intensity is dominated by shot noise fluctuations. We denote these states by $\hat{\rho}_0$. The corresponding stationary solution solves the equation
\begin{align}
\frac{1}{{\rm i}\hbar}\left[\frac{\hat{p}^2}{2m},\hat{\rho}_0\right]+\mathcal{L}_w[\hat{\sigma}^{\dag}]\hat{\rho}_0=0\,,
\end{align}
and takes the general form
\begin{align}
\hat{\rho}_0=f(\hat{p})\otimes|e\rangle\langle e|\label{inco}
\end{align}
with $f(\hat{p})$ a density operator defined over the Hilbert space of the external degrees of freedom. This density operator can be cast in terms of an analytic function of the operator $\hat p$ and thus trivially commutes with the kinetic energy term. 

We now search for stationary solutions that fulfill $X\neq0$. In particular, we require that the absolute value of the order parameter, $|X|$, is stationary. After imposing that the mean-field energy becomes constant, the condition on $|X|$ necessarily implies that the mean kinetic energy becomes constant, \textit{i.e.} $d\langle \hat{p}^2\rangle/dt=0$. 
We assume a stationary state and thus set Eq.~\eqref{energychange} to zero, with the prescription that $X=|X|e^{{\rm i}\phi(t)}$ (where $|X|>0$ and constant).
We then obtain an equation for the phase of the order parameter, which reads
\begin{align}
\frac{d\phi}{dt}=-\frac{w}{2\tan\chi}.\label{dphidt}
\end{align}
This shows that, in the regime in which the field is coherent, it oscillates in the atomic reference frame at the frequency 
\begin{equation}
\omega_w=-\frac{w}{2\tan\chi}=-\frac{\Delta w}{\kappa}\,,\label{omegaw}
\end{equation}
and thus at frequency $\omega_a+\omega_w$ in the laboratory frame. We denote the corresponding class of states by $\hat \rho_X$. These states are solution of the equation
\begin{align*}
\frac{\partial\hat{\rho}_X}{\partial t}=\frac{1}{{\rm i}\hbar}\left[-\hbar\omega_w\hat{\sigma}^{\dag}\hat{\sigma},\hat{\rho}_X\right]
\end{align*}
or alternatively
\begin{align}
\mathcal{L}_{\mathrm{mf}}[\hat{\rho}_X]\hat{\rho}_X=\frac{1}{{\rm i}\hbar}\left[-\hbar\omega_w\hat{\sigma}^{\dag}\hat{\sigma},\hat{\rho}_X\right]\,.\label{steadystate}
\end{align}
We remark that the frequency shift is proportional to the incoherent pump rate. This result has already been reported in Ref.~\cite{Tucker:2018} and was interpreted there as a signature of synchronization.

\subsection{Stability analysis}\label{stabilityanalysis}

We now investigate the stability of the stationary states that we have identified. To calculate the stability we consider small fluctuations $\delta\hat{\rho}$ about the steady-state density matrix $\hat{\rho}_{\rm st}$, namely:
\begin{align}
\hat{\rho}_1=\hat{\rho}_{\rm st}+\delta\hat{\rho}, \label{rho+deltarho}
\end{align} 
where ${\rm Tr}\{\delta\hat{\rho}\}=0$.

We first consider the class of solutions $\hat\rho_X$ with $X\neq~ 0$. For these solutions it is convenient to perform the stability analysis in the reference frame rotating with the frequency  $\omega_w$, namely, $\tilde{\rho}=e^{-{\rm i}\omega_w t\hat{\sigma}^{\dag}\hat{\sigma}}\hat{\rho}e^{{\rm i}\omega_wt\hat{\sigma}^{\dag}\hat{\sigma}}$. In this reference frame $\tilde{X}=X[\hat{\tilde{\rho}}]=e^{-{\rm i}\omega_w t}X[\hat{\rho}_1]$ and $\partial_t\hat{\tilde\rho}_X=0$. Moreover, we denote by $\tilde{\mathcal{L}}_{\mathrm{mf}}$ the transformed mean-field Lindblad operator. Substituting now Eq.~\eqref{rho+deltarho} in the mean-field master equation and dropping terms of second order in the fluctuations we get a linear equation for the time evolution of $\delta\hat{\rho}$:
\begin{align}
\frac{\partial \delta\hat{\tilde{\rho}}}{\partial t}=&\tilde{\mathcal{L}}_{\mathrm{mf}}[\hat{\tilde{\rho}}_X]\delta\hat{\tilde\rho}\label{linearVlasov}\\
&-\frac{\hbar\Lambda}{2{\rm i}\hbar\sin\chi}
\left[\left({\rm e}^{-{\rm i}\chi} \delta \tilde{X}^*\hat{\sigma}\cos(k\hat{x})+{\rm H.c.}\right),\hat{\tilde{\rho}}_X\right]\,,\nonumber
\end{align}
with $\delta \tilde{X}=X[\delta\hat{\tilde\rho}]$. Equation~\eqref{linearVlasov} is reminiscent of the linearized Vlasov equation~\cite{Campa:2009} although here it has a quantum character since it has been derived from the full quantum master equation. In order to solve Eq.~\eqref{linearVlasov} we use the Laplace transform,
\begin{align}
\mathrm{L}[f](s)=\int_{0}^\infty\,dte^{-st}f(t)\,,\label{laplace}
\end{align}
we apply it to both sides of Eq.~\eqref{linearVlasov} :
\begin{align}
s&\mathrm{L}[\delta\hat{\tilde\rho}]-\delta\hat{\tilde\rho}(0)=\tilde{\mathcal{L}}_{\mathrm{mf}}[\hat{\tilde \rho}_X]\mathrm{L}[\delta \hat{\tilde{\rho}}]\\
&+{\rm i}\frac{\Lambda}{2\sin\chi}\left({\rm e}^{-{\rm i}\chi}\mathrm{L}[\delta \tilde{X}^*][\hat{J}_1,\hat{\tilde \rho}_X]+{\rm e}^{{\rm i}\chi}\mathrm{L}[\delta \tilde{X}][\hat{J}_1^{\dag},\hat{\tilde \rho}_X]\right)\,,\nonumber
\end{align}
where we have used Eq.~\eqref{Jl} for $\ell=1$. This equation is conveniently rewritten in the form:
\begin{align}
\mathrm{L}[\delta\hat{\tilde\rho}]=&\left(s-\tilde{\mathcal{L}}_{\mathrm{mf}}\right)^{-1}\delta\hat{\tilde\rho}(0)\nonumber\\
&+{\rm i}\frac{\Lambda}{2\sin\chi}{\rm e}^{-{\rm i}\chi}\mathrm{L}[\delta \tilde{X}^*]\left(s-\tilde{\mathcal{L}}_{\mathrm{mf}}\right)^{-1}[\hat{J}_1,\hat{\tilde\rho}_X]\nonumber\\
&+{\rm i}\frac{\Lambda}{2\sin\chi}{\rm e}^{{\rm i}\chi}\mathrm{L}[\delta \tilde{X}]\left(s-\tilde{\mathcal{L}}_{\mathrm{mf}}\right)^{-1}[\hat{J}_1^{\dag},\hat{\tilde\rho}_X]\,.
\end{align}
We now multiply by $\hat{J}_1$ and $\hat{J}_1^{\dag}$, respectively, and take the trace. Using the fact that $\mathrm{L}[\delta \tilde{X}]={\rm Tr}\{\mathrm{L}[\delta\hat{\tilde\rho}]J_1\}$ we obtain two coupled equations, which can be cast into matrix form
\begin{align}
\label{D:matrix}
{\bf D}(s)\begin{pmatrix}
\mathrm{L}[\delta \tilde{X}](s)\\
\mathrm{L}[\delta \tilde{X}^*](s)
\end{pmatrix}={\bf b}(s).
\end{align}
Here ${\bf D}(s)=1_2+{\bf C}(s)$ where $1_2$ is the $2\times2$ identity matrix, ${\bf C}$ is defined by
\begin{align}
{\bf C}=\begin{pmatrix}
C_{11}&C_{12}\\
C_{21}&C_{22}
\end{pmatrix},\label{CMatrix}
\end{align}
and the vector ${\bf b}$ has the form
\begin{align}
{\bf b}=\begin{pmatrix}
b_{1}\\
b_{2}
\end{pmatrix}.
\end{align}
The corresponding entries are given in detail in Appendix~\ref{App:C}.

In the form of Eq.~\eqref{D:matrix} the stability analysis consists now of determining the first order poles $\tilde{\gamma}$ of ${\bf D}(s)^{-1}{\bf b}(s)$. The poles, in fact, determine the short-time dynamics of  $\delta \tilde{X}(t), \delta \tilde{X}^*(t)$ according to $\delta \tilde{X}(t)\sim e^{\tilde{\gamma} t}$, with $\delta X(t)\sim e^{\gamma t}$ and 
$$\gamma=\tilde{\gamma}+{\rm i}\omega_w$$
is the pole in the reference frame rotating at the atomic frequency. 
The stationary state is unstable if there exists a solution with $\mathrm{Re}(\tilde{\gamma})>0$. On the contrary, it is stable if the real parts of all of the eigenvalues are negative.

We can now use  this machinery to calculate the stability of the state $\hat{\rho}_0$ (see Eq.~\eqref{inco}) assuming that the momentum distribution is thermal,
 \begin{align}
f(\hat{p})=\frac{1}{Z}\exp\left(-\beta\frac{\hat{p}^2}{2m}\right),\label{homGaussian}
\end{align}
with $Z=\sqrt{2m\pi/\beta}$. In this case $\tilde{\mathcal{L}}_{\mathrm{mf}}$ describes the evolution of free particles and thus an instability can only be due to the singularities of the matrix ${\bf D}$.
We thus analyze 
\begin{align}
D(s)=\det({\bf D}(s))=0\,.\label{dispersion relation}
\end{align}
We note that Eq.~\eqref{dispersion relation} gives the dispersion relation in the stable limit. In Appendix \ref{App:C} we show that the dispersion relation takes the form 
\begin{align}
\frac{y}{\sqrt{\pi\sigma_{\beta}^2}}\int_{-\infty}^{\infty} du\frac{\exp\left(-\frac{u^2}{\sigma_\beta^2}\right)}{y^2+u^2}=
4{\rm e}^{-{\rm i}\chi}\sin\chi
\label{hogaussdis}
\end{align}
with $y=(s-{\rm i e}^{{\rm i}\chi}w/(2\sin\chi)-{\rm i}\omega_R)/\Lambda$ and $\sigma_\beta^2=\bar\beta/\beta$, and we have introduced the temperature scale
$$\bar\beta^{-1}=\frac{\hbar \Lambda^2}{2\omega_R}\,,$$ 
which depends on the ratio between the superradiant linewidth $\Lambda$ and the recoil frequency $\omega_R$. 
The physical meaning of this temperature scale will be discussed in the following. We note here that the solution of Eq.~\eqref{hogaussdis} depends solely on $\sigma_\beta$ and $\chi$. 

We now denote by $\gamma'$ a solution of Eq.~\eqref{hogaussdis}. Thus the corresponding value of the pole $\tilde \gamma$ takes the form 
\begin{align}
\tilde\gamma=\gamma'-\frac{w}{2}+{\rm i}\left(\omega_R-\omega_w\right)\label{gammagammaprime}.
\end{align}
We first analyze the form of the solution in the limit $\sigma_\beta\ll 1$. In this case we can solve Eq.~\eqref{hogaussdis} to first-order in $\sigma_\beta$ and obtain 
\begin{align}
\gamma=&\frac{\Lambda}{4}-\frac{w}{2}-4\Lambda\sin^2\chi\frac{\bar\beta}{\beta}\\
&+{\rm i}\left(\omega_R-\frac{\Lambda}{4\tan\chi}-4\sin\chi\cos\chi\frac{\bar\beta}{\beta}\right).\label{gammap=0}
\end{align}
Let us first consider $\beta\to\infty$, or equivalently $T=0$: This is the limit in which we neglect the thermal fluctuations. The stationary solution $\hat{\rho}_0$ is then stable for $w>\Lambda/2$, otherwise it becomes unstable. For $w<\Lambda/2$, in particular, small fluctuations are amplified and lead to the growth of $X$ and thus of the intracavity field. We note that this bound coincides with the one predicted by a semiclassical calculation for this setup and a homogeneous medium \cite{Jaeger:2017}. Equation~\eqref{gammap=0} also shows that small thermal fluctuations tend to stabilize the state $\hat\rho_0$, shifting the bound to lower values than $\Lambda/2$.

\begin{figure}[h!]
	\center \includegraphics[width=0.9\linewidth]{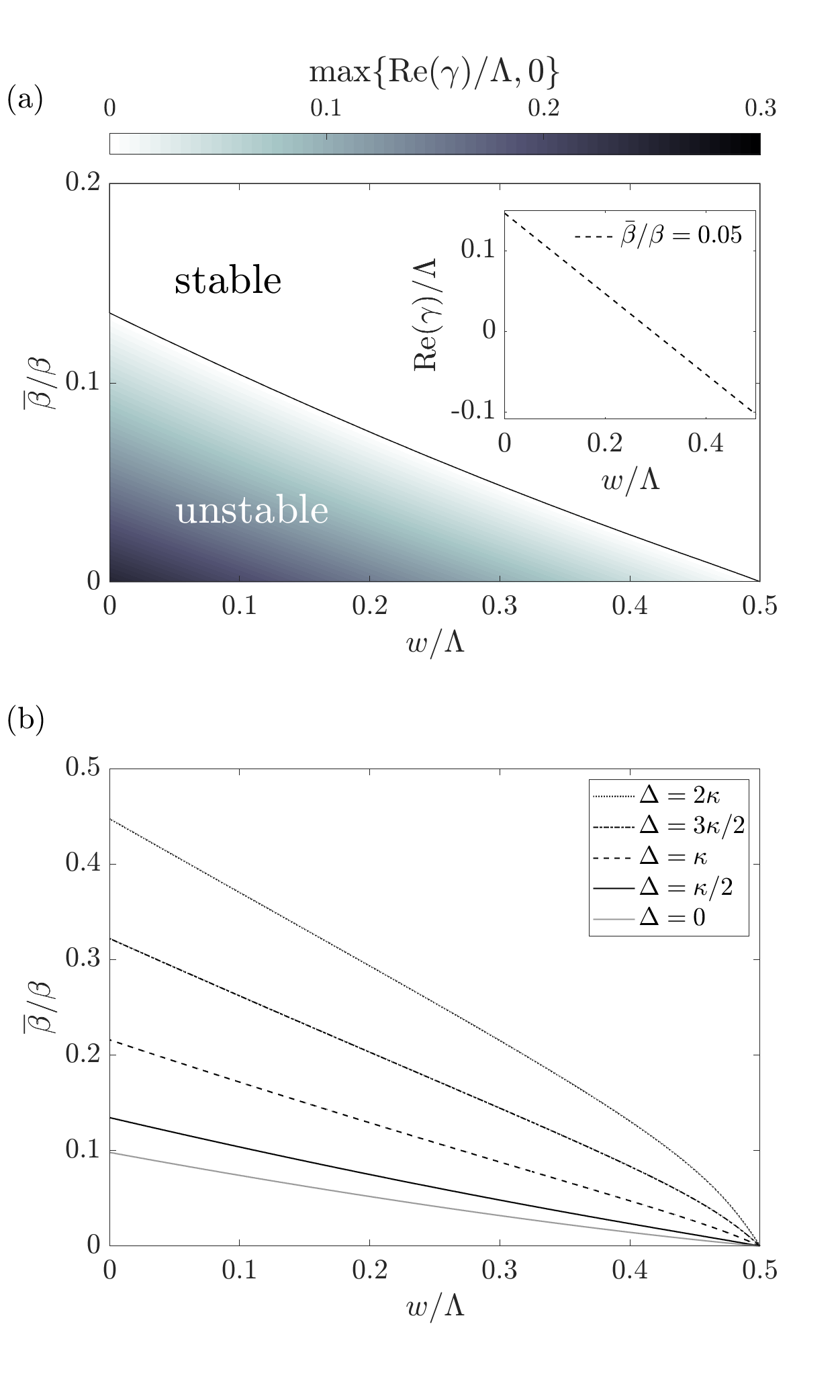}
	\caption{(a) Diagram in the $\beta-w$ plane of the stability of the state $\hat{\rho}_0$: when $\hat{\rho}_0$ is stable there is no superradiant emission. The contour plot gives the maximal real value of the solutions $\gamma$ of Eq. \eqref{hogaussdis}: when these are positive (below the black solid line) the intracavity field grows exponentially with rate ${\rm Re}(\gamma)$. The rate $\gamma$ and the incoherent pump rate $w$ are reported in units of $\Lambda$, the inverse temperature $\beta$ in units of $\bar{\beta}$. The phase diagram has been calculated for $\Delta=\kappa/2$, corresponding to $\tan\chi=1$. The inset displays ${\rm Re}(\gamma)$ as a function of $w$ for $\bar\beta/\beta=0.05$. Subplot~(b) shows the curve $w(\beta)$ separating the region of stability from the one of superradiant emission and for different positive values of $\Delta$ (see inbox).\label{stability}}
\end{figure}

Figure~\ref{stability} displays a contour plot of the real value of the solutions $\gamma$ of Eq.~\eqref{hogaussdis}, which are found by numerically solving the equations. Among all solutions, we plot the ones that are characterized by the maximal real value for given $\sigma_\beta$ and $\chi$. The solutions are determined as a function of $\beta$ and of $w$, keeping constant all other physical parameters (thus also $\Lambda$ and $\bar\beta$). The black line divides the region where the state $\hat\rho_0$ is unstable from the region where it is stable, thus corresponding to the solutions of Eq.~\eqref{hogaussdis} for which the real value of the eigenvalues $\gamma$ vanish, and fit our analytical estimate, Eq.~\eqref{gammap=0}, for $\bar\beta/\beta\to 0$. The instability corresponds to the fast growth of the intracavity field, and thus to a superradiant pulse emitted on a timescale of the order of $1/\Lambda$. The plot also shows that there is a critical temperature $T_c$ above which superradiant emission is suppressed. This temperature is given by
$$k_BT_c\approx 0.1{\bar \beta}^{-1}=0.1\frac{\hbar \Lambda^2}{2\omega_R}\,,$$
where $k_B$ is the Boltzmann constant. For $T>T_c$ thermal fluctuations suppress the quantum interference process that lies at the basis of the superradiance mechanism. 

The value of $T_c$ depends also on the detuning $\Delta$, as is visible from Fig.~\ref{stability}(b). This Figure displays the line $w(\beta)$, below which there is superradiant emission, as a function of $\Delta$ and shows that the temperature interval for which the intracavity field initially grows increases monotonously with $\Delta$. All curves tend to the same value for $\beta\to \infty$ ($T\to 0$).

\subsection{Numerical analysis}\label{relaxationthermalstate}

In this section we numerically integrate Eq.~\eqref{MEq:MF} and analyze the dynamics of the order parameter $X(t)$, Eq.~\eqref{mforderparameter}, when the initial state is given by Eq.~\eqref{inco} and Eq.~\eqref{homGaussian}. The simulations reported in this section are performed taking $k_BT=20\hbar \omega_R$, such that the dynamics is in the semiclassical regime. Moreover, we take $\Lambda=40\omega_R$, and therefore $k_B T_c\approx 0.1 \hbar \Lambda^2/(2\omega_R)=80 \hbar \omega_R$. The initial temperature is taken to be below the threshold and warrants that the system evolves towards superradiant emission provided that $w<0.5\Lambda$ for $\Delta=\kappa/2$, see Fig.~\ref{stability}. The numerical simulation is performed on a grid of momentum states in the interval $[-p_{\mathrm{max}},p_{\mathrm{max}}]$ with $p_{\mathrm{max}}=16\hbar k$ and discrete steps $\Delta p=\hbar k/10$. Here and in the rest of this work the integration interval and grid are chosen after checking the convergence of the dynamics of the order parameter.

\begin{figure}[h!]
	\center \includegraphics[width=0.8\linewidth]{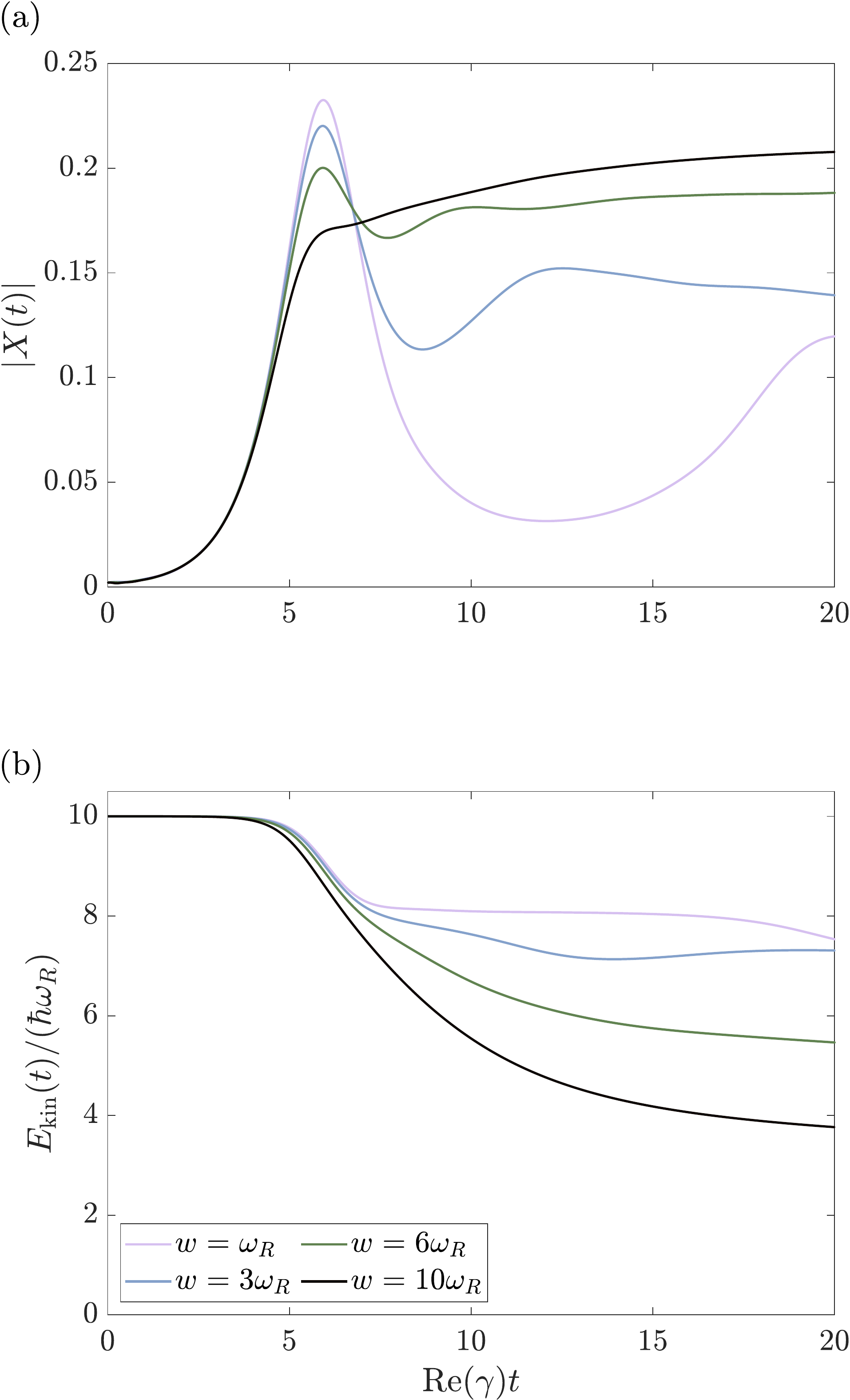}
	\caption{ Dynamics of (a) the absolute value of the mean-field order parameter $|X(t)|$ and (b) of the mean kinetic energy $\langle \hat{p}^2/(2m)\rangle$ (in units of the recoil energy $\hbar\omega_R$) for different values of the pump $w=1,3,6,10\omega_R$, see the legend in the inset of subplot (b). The parameters and details of the simulation are given in the text, the time is reported in units of $\mathrm{Re}(\gamma)^{-1}$ where $\gamma$ has been determined using Eq.~\eqref{hogaussdis} and depends on~$w$. \label{dyn}}
\end{figure}

\begin{figure}[h!]
	\center \includegraphics[width=0.8\linewidth]{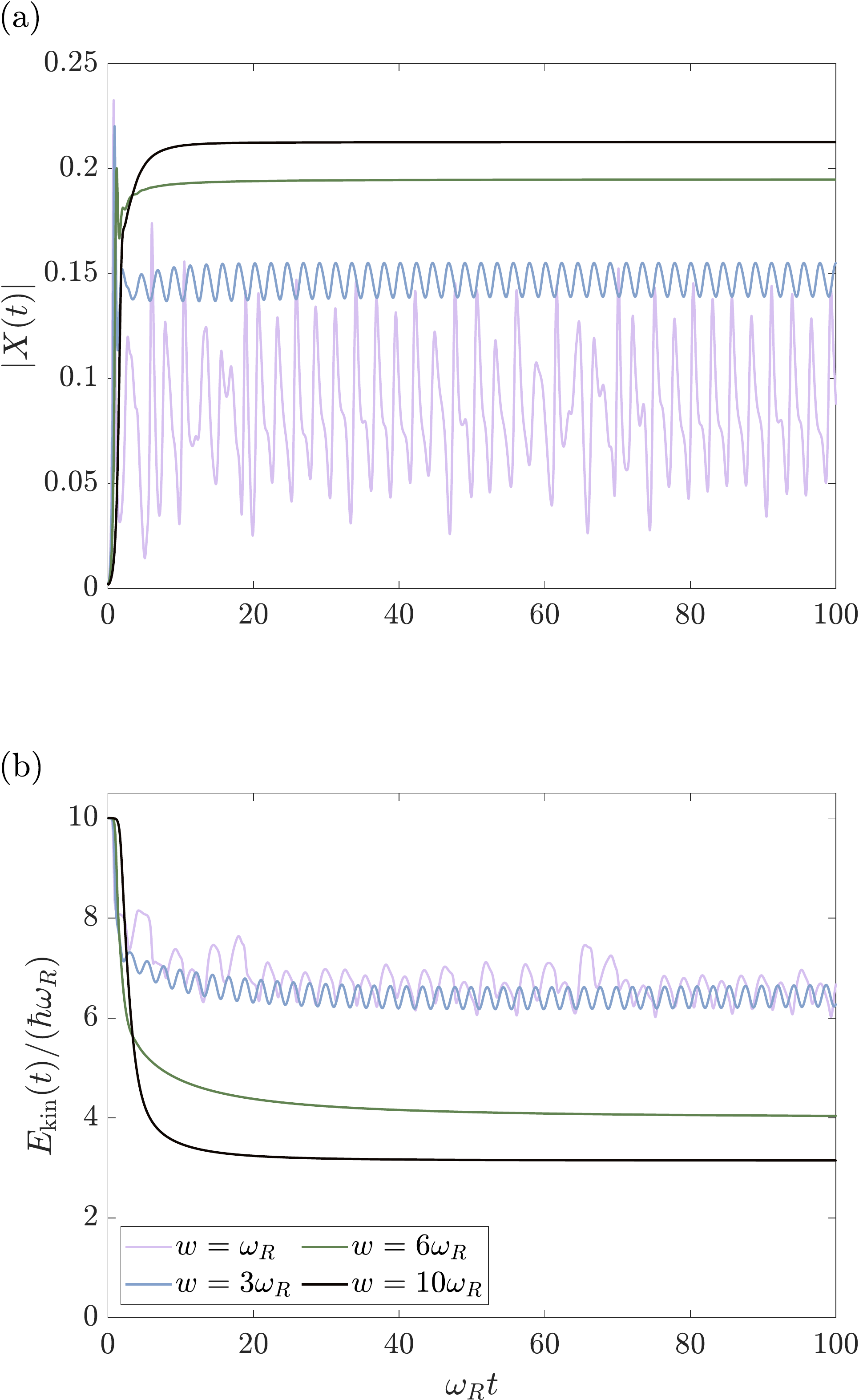}
	\caption{Dynamics of (a) the absolute value of the mean-field order parameter $|X(t)|$ and (b) of the mean kinetic energy $\langle \hat{p}^2/(2m)\rangle$ (in units of the recoil energy $\hbar\omega_R$) for different values of the pump $w$, same as Fig.~\ref{dyn}. The time is now reported in units of $\omega_R^{-1}$ and the dynamics is integrated more then ten times longer.\label{dyn:2}}
\end{figure}

Figure~\ref{dyn}(a) displays the dynamics of $|X(t)|$ for different values of $w$ (in the interval where we expect superradiant emission). All curves show an initial exponential growth, after which they exhibit different behaviors (which we discuss later when analyzing the results displayed in Fig.~\ref{dyn:2}). We note that the time has been rescaled by the real part of the exponent $\gamma$, which we extract from Eq.~\eqref{hogaussdis}, and thus depends on $w$. This rescaling allows us to verify the correctness of our stability analysis for short times: In fact, all curves collapse on a single exponential growth for short times.
Figure~\ref{dyn}(b) displays the corresponding dynamics of the mean kinetic energy. The latter slowly decreases over the timescale in which $|X|$ exponentially grows. It then decreases with larger rates, but following a behavior that depends on $w$ and that can be understood analyzing $|X(t)|$ and using Eq.~\eqref{kineticdyn}. 

For short timescales Eq.~\eqref{kineticdyn} allows us to determine the conditions on the parameter that lead to the initial decrease of the kinetic energy. For this purpose we use $|X(t)|\approx|\delta X(0)|{\rm e}^{\mathrm{Re}(\gamma) t}$ and Eq.~\eqref{gammap=0}, which is valid for $\beta\gg\bar\beta$, hence for very low temperatures. We obtain 
\begin{align}
\frac{d}{dt}\frac{\langle \hat{p}^2\rangle}{2m}
\approx&\hbar\Lambda|\delta X|^2\left[\omega_R-8\Lambda\sin\chi\cos\chi\sigma_\beta^2\right]\,,
\label{Kin:stability}
\end{align}
and thus the necessary condition for the initial decay of the kinetic energy is that $\Delta>0$ (for which $\sin\chi\cos\chi>0$). We then get an upper bound for $\sigma_\beta^2$, namely, $\sigma_\beta^2<~\omega_R/(4\Lambda\sin(2\chi))$, which shall be considered when it is much smaller then unity, namely, when $4\Lambda\sin(2\chi)\gg \omega_R$. 

The dynamics of the order parameter and mean kinetic energy for longer timescales is shown in Fig.~\ref{dyn}(a) and Fig.~\ref{dyn}(b). For all considered values of $w$ the time-averaged values of the order parameter $|X|$ and of the kinetic energy reach a stationary value that is different from zero and that monotonically increases and decreases, respectively, with $w$. Moreover, for $w=1,3\omega_R$ we observe oscillations about the stationary value, whose amplitude become significantly larger for $w=\omega_R$ and which are evidently polychromatic. 

\begin{figure}[h!]
	\center \includegraphics[width=0.9\linewidth]{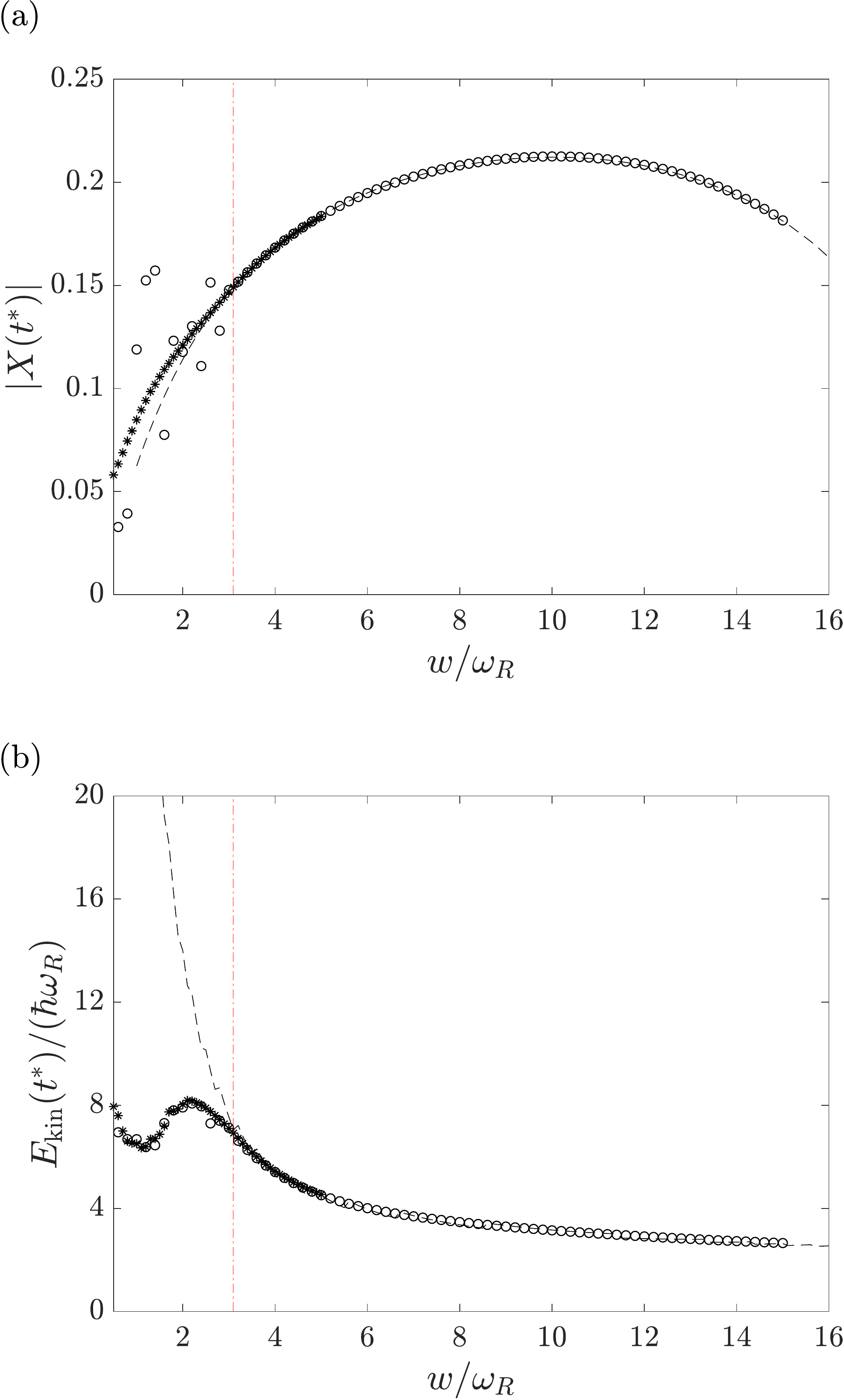}\\
	\caption{The time-averaged mean values (a) $|X|_{\rm st}$ and (b) $\langle \hat{p}^2\rangle_{\rm st}/(2m)$ as a function of $w$, as obtained by numerically integrating Eq.~\eqref{MEq:MF} for the initial state in Eq.~\eqref{inco} (star symbols). The circles correspond to the numerical values of (a) $|X(t)|$ and (b) of the kinetic energy at time $t=400\omega_R^{-1}$. The dashed line gives the stationary solution of Eq.~\eqref{steadystate}. The vertical dotted line indicates the threshold value $w_c\approx =3.1\omega_R$, below which the three curves do not overlap, and which agrees with the value we find by solving the dispersion relation (Eq.~\eqref{dispersion relation}) for the stationary state. The other parameters are given in the text. \label{Steady:thermal}}
\end{figure}

We numerically determine the time-averaged mean values by evaluating the quantity
\begin{align}
|X|_{\rm st}=&\frac{1}{t_{\mathrm{sim}}}\int_{0}^{t_{\mathrm{sim}}}\,dt|X|(t)\,,\label{Xa}\\
\langle \hat{p}^2\rangle_{\rm st}=&\frac{1}{t_{\mathrm{sim}}}\int_{0}^{t_{\mathrm{sim}}}\,dt\langle \hat{p}^2\rangle(t)\,,\label{pa}
\end{align}
where $t_{\mathrm{sim}}$ is the time of the simulation, which is taken to be $t_{\mathrm{sim}}=1200\omega_R^{-1}$ after checking the convergence of the results over a sufficiently large sample of simulation intervals. Their behavior as a function of $w$ is displayed in Fig.~\ref{Steady:thermal}(a) and (b). We compare them with the corresponding behavior of $|X(t)|$  and $\langle \hat{p}(t)^2\rangle/(2m)$ at a specific instant of time, $t=400\omega_{R}^{-1}$ (circles), as well as with the solutions we extract by analytically solving Eq.~\eqref{steadystate} (dashed line). This comparison allows us to identify a threshold value of the pump strength $w_c$ above which the three values agree, and below which there are evidently discrepancies between the three solutions. This threshold value agrees with the one we find by solving Eq.~\eqref{dispersion relation} and is estimated to be $w_c\approx 3.1\,\omega_R$. Remarkably, for $w<w_c$, the time-averaged mean kinetic energy is well below the stationary value determined analytically. 

\begin{figure}[h!]
	\center \includegraphics[width=1\linewidth]{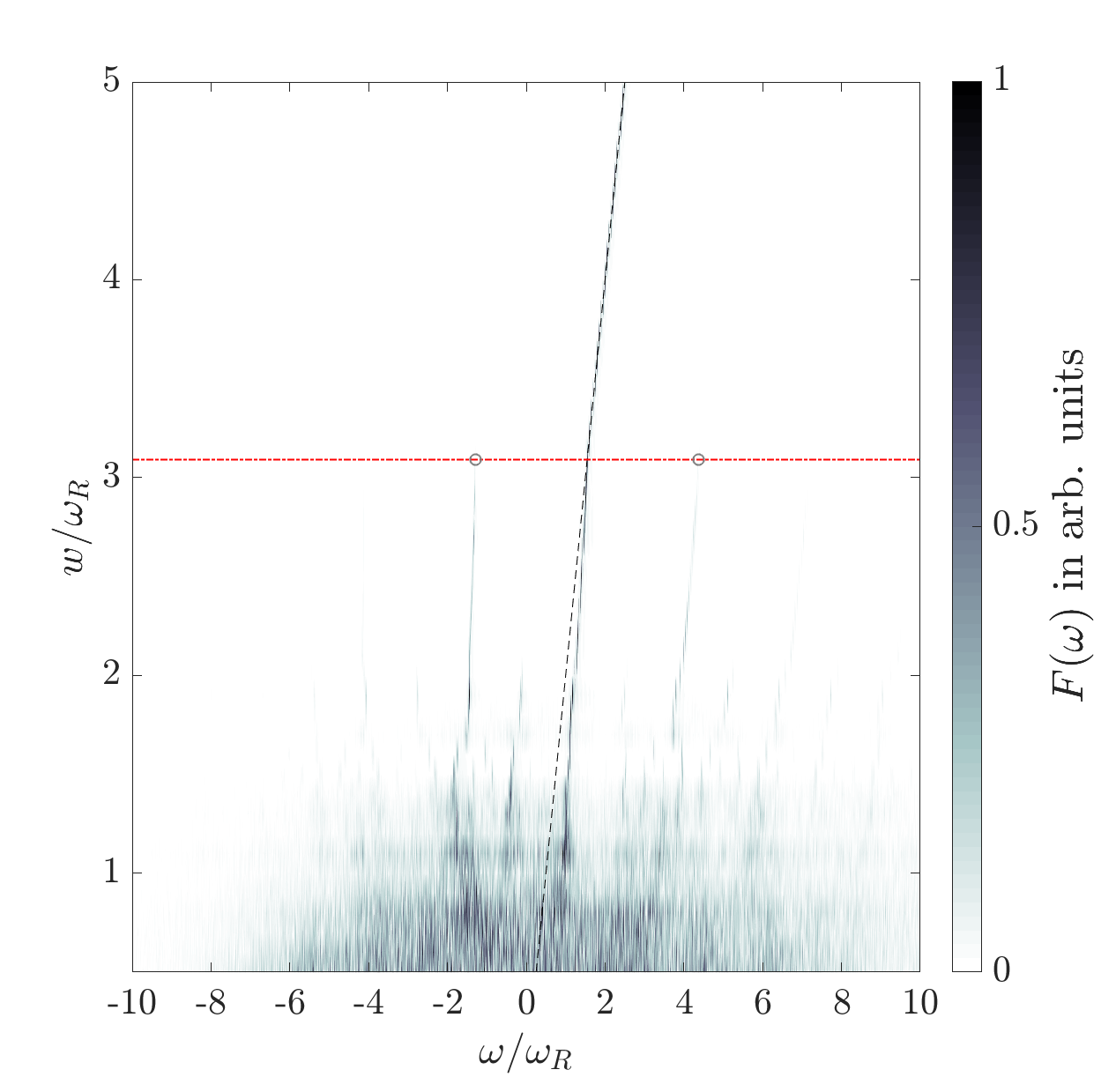}\\
	\caption{Contour plot of the spectrum $F(\omega)$, Eq.~\eqref{fourier}, as a function of the pump rate $w$ and of the frequency $\omega$ (both in units of $\omega_R$). For every value of $w$ we normalize $F(\omega)$ with its maximum value. The dashed black line corresponds to the frequency $\Delta w/\kappa$, Eq.~\eqref{dphidt}. The dashed-dotted gray horizontal line gives the value of $w_c$. The circles correspond to the imaginary part of the solutions $\gamma$ of Eq.~\eqref{dispersion relation}. The other parameters are given in the text.\label{spectrumthermal}}
\end{figure}

In order to gain insight into the oscillatory behavior at $w<w_c$ we calculate the spectrum of $X(t)$, which we define as 
\begin{align}
F(\omega)=\left|\int_{0}^{t_\mathrm{sim}}\,dte^{{\rm i}\omega t}X(t)\right|\,.\label{fourier}
\end{align}
This quantity squared is proportional to the power spectrum at the cavity output and can thus be accessed by means of photodetection. 
Figure~\ref{spectrumthermal} displays the contour plot of $F(\omega)$ in the $w-\omega$ plane. For clarity we indicate the value $w=w_c$ with the horizontal dashed-dotted line. For $w>w_c$ it exhibits a single frequency peak at the frequency $\Delta w/\kappa$, in agreement with the result of Eq.~\eqref{dphidt}. At $w=w_c$ two sidebands appear. Their frequency agrees with the imaginary parts of the solutions $\gamma$ of the dispersion relation~\eqref{dispersion relation}. As $w$ decreases below $w_c$ an increasing number of sidebands appears, until the spectrum becomes nearly continuous. This dynamics presents the characteristic features of classical chaos. Despite the fact that the mean kinetic energy seems to reach a stable value, on the other hand the momentum distribution has long tails. We will discuss this behavior more in detail in the next section.

\section{Superradiance in a Bose-Einstein condensate}\label{Sec:BEC}

We now study the effect of quantum fluctuations on stationary superradiance. For this purpose we focus on the regime where the superradiant linewidth $\Lambda$ takes values comparable to the recoil frequency $\omega_R$ and assume that the atoms form initially an ideal Bose-Einstein condensed (BEC) gas and occupy the state at momentum $p=0$:
\begin{equation}
\label{Eq:BEC}
\hat{\rho}_1(0)=|e\rangle\langle e|\otimes |p=0\rangle\langle p=0|\,.
\end{equation}
Since the mechanical effects of light result from the absorption and emission of cavity photons with linear momentum $\pm \hbar k$,  the atomic momentum can take only the values $|\Psi_0\rangle=|0\rangle$ (the BEC) and $|\Psi_n\rangle=(|n \hbar k\rangle+|-n \hbar k\rangle)/\sqrt{2}$ $(n=1,2,\ldots)$. These states have kinetic energy $E_{{\rm kin},n}=n^2\hbar\omega_R$. On this grid we can numerically solve Eq.~\eqref{steadystate} as a function of $w$ and $\Lambda$ and for the momentum cutoff $p_{\rm max}=15\hbar k$. For this purpose we use a seed $X>0$, determine the corresponding stationary value $\hat{\rho}_0$, and use it  to determine the new value of $X=\mathrm{Tr}(\hat{\sigma}\cos(k\hat{x})\hat{\rho}_0)$. We iterate this procedure until $\hat{\rho}_0$ and $X$ have converged. We then determine the stability of the solution using the dispersion relation of Eq.~\eqref{dispersion relation}. 

We identify four phases. One phase corresponds to the situation where the initial state is stable. This is found for $w>\Lambda/2$ and corresponds to the case in which there is no superradiant emission. This result has already been discussed in Sec.~\ref{sec:3} and reported in Fig.~\ref{stability}. The other three phases are found for $w<\Lambda/2$. Here, in absence of the optomechanical coupling one expects stationary superradiance over the whole interval and for any value of $\Lambda$. When taking into account the mechanical effects of light on the dynamics, we recover stationary superradiance, corresponding to the stable solution $X\neq 0$, for $w<\Lambda/2$ provided that both $w$ as well as $\Lambda$ exceed threshold values of the order of the recoil frequency. Below these threshold values the phase can be characterized by a stable stationary solution at $X=0$, which we denote as an incoherent phase, or by the absence of stable solutions. When there is no stable solution, both the incoherent and the superradiant states are unstable. We denote this phase by ``chaotic" as will become clearer when analyzing the time-dependent dynamics. Figure~\ref{phasediagram} displays the phase diagram in the $w-\Lambda$ plane, the phases are labeled by the properties of the order parameter $X$, and thus of the field at the cavity output. 

Below we discuss the behavior across the direct transition from the incoherent to the coherent phase and from the coherent to the chaotic phase. We remind the reader that the transition is driven-dissipative, moreover, it is dynamical, depends on the initial state, and occurs over a timescale of the order of the superradiant decay time. After this timescale the atoms organize so as to emit either coherent, incoherent, or chaotic light.

\subsection{Transition from incoherent to coherent phase.}\label{PathA}

\begin{figure}[h!]
	\center \includegraphics[width=0.9\linewidth]{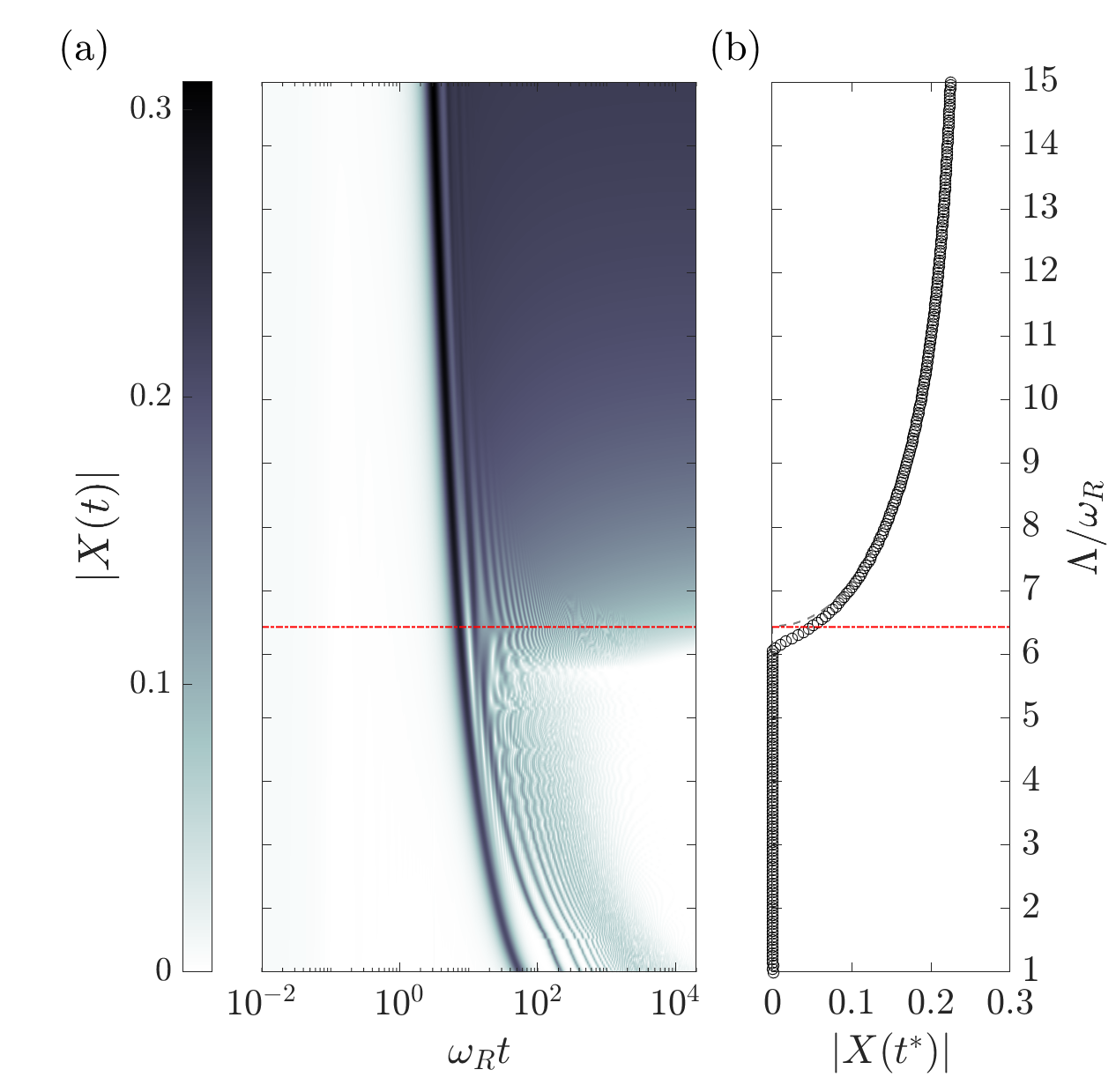}\\
	\caption{(a) Contour plot of $|X(t)|$ as a function of the time $t$ (in units of $\omega_R^{-1}$) and of $\Lambda$ (in units of $\omega_R$) for $w=\Lambda/4$ and $\Delta=\kappa/2$. The dynamics of $|X(t)|$ is calculated by numerically integrating Eq.~\eqref{MEq:MF} with the initial state of Eq. \eqref{Eq:BEC} over a grid of momenta with a cut-off at $p_{\mathrm{max}}=15\hbar k$.  (b) The asymptotic value of $|X(t)|$ is displayed as a function of $\Lambda$ (in units of $\omega_R$). The dashed line is the solution of Eq.~\eqref{steadystate}, the circles correspond to $X(t^*)$, as obtained by integrating Eq. \eqref{MEq:MF} until $t^*=4\times 10^4\omega_R^{-1}$. The red dashed-dotted line in both subplots indicates the critical value $\Lambda_c$ at which the steady state solution of Eq.~\eqref{steadystate} predicts a transition from the incoherent to the coherent phase. \label{dynbreakdown:X}}
\end{figure}

The phase diagram in Fig.~\ref{phasediagram} displays a parameter regime where the solution of Eq.~\eqref{steadystate} predicts a direct transition between the incoherent and the coherent phase across the curve $\Lambda_c(w)$. In order to analyze in detail the behavior we consider a specific parameter variation that crosses the expected phase transition line at the value $\Lambda_c=\min_w\Lambda_c(w)$, and specifically vary $w$ and $\Lambda$ together with the constraint $w=\Lambda/4$. The dynamics of $|X(t)|$ along this line is summarized in the contour plot of Fig.~\ref{dynbreakdown:X}(a) in the $\Lambda-t$ plane. Here, the timescale of the first superradiant emission is visible in the first maximum of $|X(t)|$ as a function of time. Our stability analysis predicts that this timescale depends on $\Lambda$ according to an exponential function and the position in time of this maximum is in agreement with our analytical prediction. The stationary behavior becomes visible after this maximum: below a certain value of $\Lambda$ the signal exhibits damped oscillations until it becomes zero. Above this value, after few transient oscillations it tends to a constant value different from zero, which increases with $\Lambda$. The boundary between these two regimes can be drawn according to various alternate criteria, which all locate the transition point to be around $6\omega_R$. In the figures we show the boundary (red line) we extract from the solution of Eq.~\eqref{steadystate}. 
 
\begin{figure}[h!]
	\center \includegraphics[width=1\linewidth]{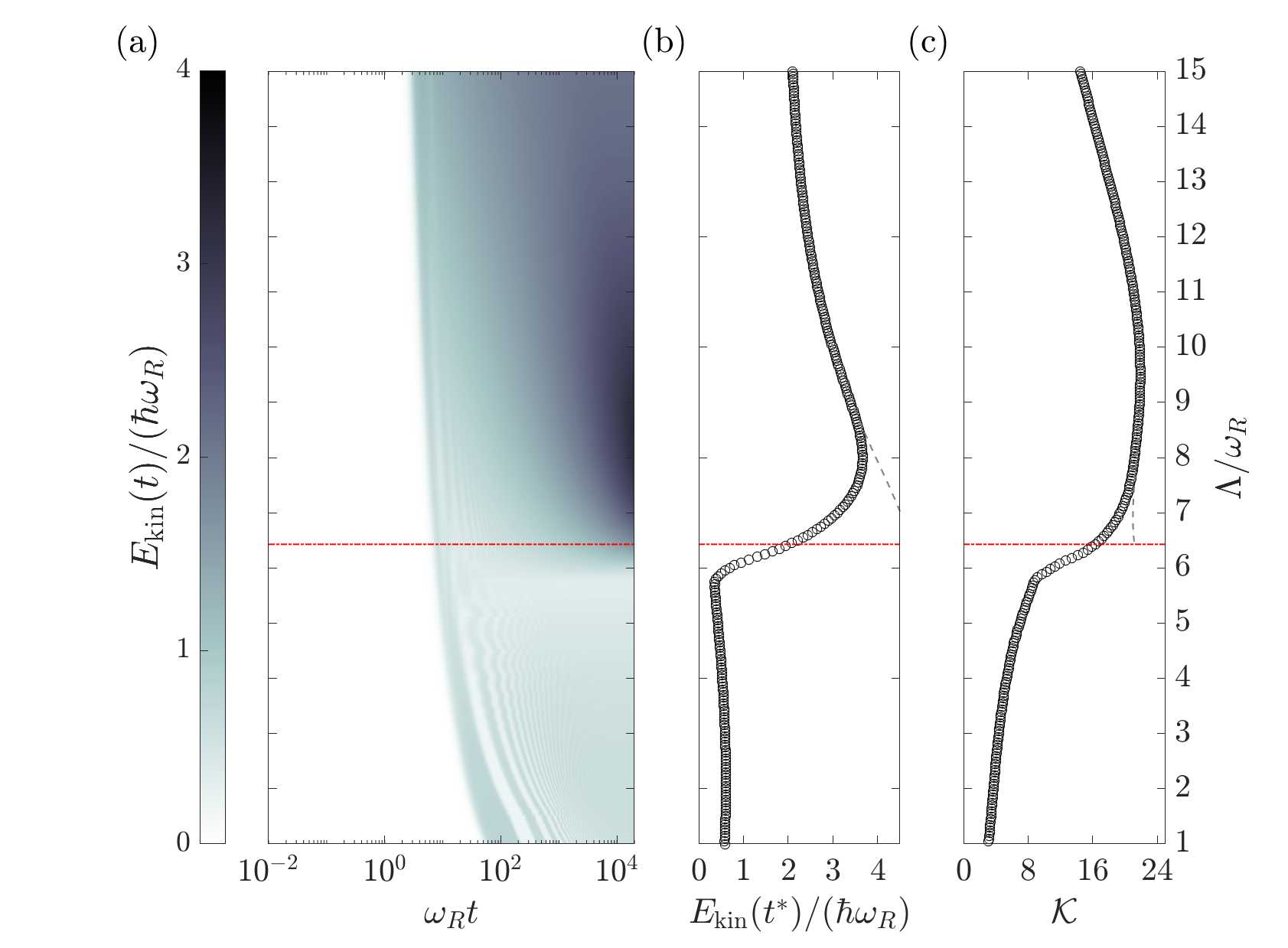}\\
	\caption{Subplots (a) and (b) are the same as in Fig. \ref{dynbreakdown:X} but for the kinetic energy $E_{\rm kin}=~\langle \hat{p}^2\rangle/(2m)$ (in units of the recoil energy $\hbar\omega_R$). Subplot (c) shows the kurtosis $\mathcal{K}(t)$ (Eq.~\eqref{Kurtosis}) at $t=t^*$ (circles) for different values of $\Lambda$ in units of $\omega_R$. The dashed line displays the corresponding value of the kurtosis, Eq.~\eqref{steadystate}. \label{dynbreakdown:Kin}}
\end{figure}

Figure~\ref{dynbreakdown:X}(b) displays the asymptotic value of $|X(t)|$ as a function of $\Lambda$ along the curve $w=\Lambda/4$. The asymptotic value is here given by the solution of Eq.~\eqref{steadystate} (dashed line) and is compared with the value of $|X(t^*)|$ which we obtain after integrating the master equation \eqref{MEq:MF} for a sufficiently long time $t^*= 4\times 10^4\omega_R^{-1}$ (circles). Both results are in good agreement, except in the vicinity of the critical point $\Lambda\approx \Lambda_c$. Here, we observe discrepancies that might be due to the finite integration time $t^*$. We have verified, in fact, that the transition point $\Lambda_c^*$, extracted from the behavior of $|X(t^*)|$, approaches $\Lambda_c$ as $t^*$ is increased.  

The corresponding behavior of the mean kinetic energy is shown in Fig.~\ref{dynbreakdown:Kin}(a) and Fig.~\ref{dynbreakdown:Kin}(b). As is visible in subplot (a), the first superradiant emission is associated with a net increase of the kinetic energy. This result is consistent with the stability analysis of Eq.~\eqref{Kin:stability}, which predicts a positive slope for $\sigma_\beta=0$. After this transient, in the incoherent phase the kinetic energy undergoes damped oscillations as a function of time until it reaches a stationary value, which is smaller than one recoil. On the contrary, in the coherent phase the oscillations seem overdamped and the stationary value is of the order of several recoil energies. Subplot~(b) displays the kinetic energy from the iterative solution of Eq.~\eqref{steadystate} and from the numerical value at $t=t^*$, both as a function of $\Lambda$. As for the order parameter, the two values of the kinetic energy are in good agreement expect for the small region close to $\Lambda_c$. This plot evidentiates that in the coherent phase the kinetic energy is significantly larger than in the incoherent phase, and it seems to undergo a discontinuous jump at the critical point. 

It is interesting to analyze the momentum distribution more in detail. For this purpose we analyze the kurtosis, namely the ratio
\begin{align}
\mathcal{K}(t)=\frac{\langle\hat{p}^4(t)\rangle}{\langle\hat{p}^2(t)\rangle^2}\label{Kurtosis}\,
\end{align}
that lets us distinguish between Gaussian ($\mathcal{K}=3$), short-tailed ($\mathcal{K}<3$), or long-tailed ($\mathcal{K}>3$) momentum distributions.
The kurtosis is displayed in Fig. \ref{dynbreakdown:Kin}(c) at $t=t^*$, corresponding to the mean kinetic energy of subplot (b). For all considered values of $\Lambda/\omega_R$ the kurtosis is larger than 3, indicating that the distribution is non-Gaussian and exhibits long tails.

It is important to point out that the increase of the kinetic energy is associated with the fact that the atomic ensemble has built up a matter wave grating which diffracts the light. In particular, for $\Lambda<\Lambda_c$ the dynamics is accompanied by the formation of a statistical mixture of states $\ket{e,\Psi_{2n}}$ and  $|e,\Psi_{2n+1}\rangle$, which dephases the macroscopic dipole and leads to suppression of the superradiant emission. For $\Lambda > \Lambda_c$, on the other hand, the field oscillates about a finite asymptotic value and the atoms form a stable spatial pattern. The corresponding state minimizes the entropy and exhibits entanglement between internal and external degrees of freedom \cite{Jaeger:2019}. This entanglement is a signature of the locking between the dynamics of external and internal degrees of freedom, which maximizes the value of the order parameter $X$.  

The picture in terms of discrete energy states $\{|e,\Psi_n\rangle,|g,\Psi_n\rangle\}$ allows one to understand the origin of the critical value $\Lambda_c$. In fact, the buildup of a stable pattern requires that coherent transitions are driven between the pair of states $|e,\Psi_n\rangle\to |e,\Psi_{n+2}\rangle$ (recall that $\Delta\gg\Lambda$). We estimate the dividing line separating the incoherent from the coherent phase by determining where the coherent transition amplitude $\mathcal{T}$ coupling the states~$\ket{e,p=0}$ and~$\ket{e,p=\pm 2\hbar k}$ is equal to their energy offset~$4\omega_R$, i.e., $\mathcal{T}\sim 4\omega_R$, so as to allow those states to be populated. Using Eq.~\eqref{MEq:MF} and the value of the order parameter which we find using the semiclassical theory of Ref. \cite{Jaeger:2017}, $|X|^2\sim w/(2\Lambda)$ , we find $$\mathcal{T}\sim (\Lambda |X|)^2/(w^2+\omega_R^2) 4\omega_R\,,$$ leading to the estimate $\Lambda_c \sim  4\omega_R$, which qualitatively agrees with our numerical result. 
 
\subsection{Transition from coherent to chaotic phase}\label{PathB}

The ``chaotic" phase of the diagram in Fig.~\ref{phasediagram} indicates that there is a lower bound to the pump rate $w$, below which there is no stationary superradiance. We remark that this lower boundary results from the optomechanical coupling. In fact, in the absence of coupling with the motion stationary superradiance is expected for $w>\Gamma_c=\Lambda/N$ (and in our mean field treatment  $\Gamma_c\to 0$). For a finite number $N$ of atoms, the behaviors we discuss in what follows can be observed provided that $\Gamma_c\ll \omega_R$. 

\begin{figure}[h!]
    \center \includegraphics[width=0.9\linewidth]{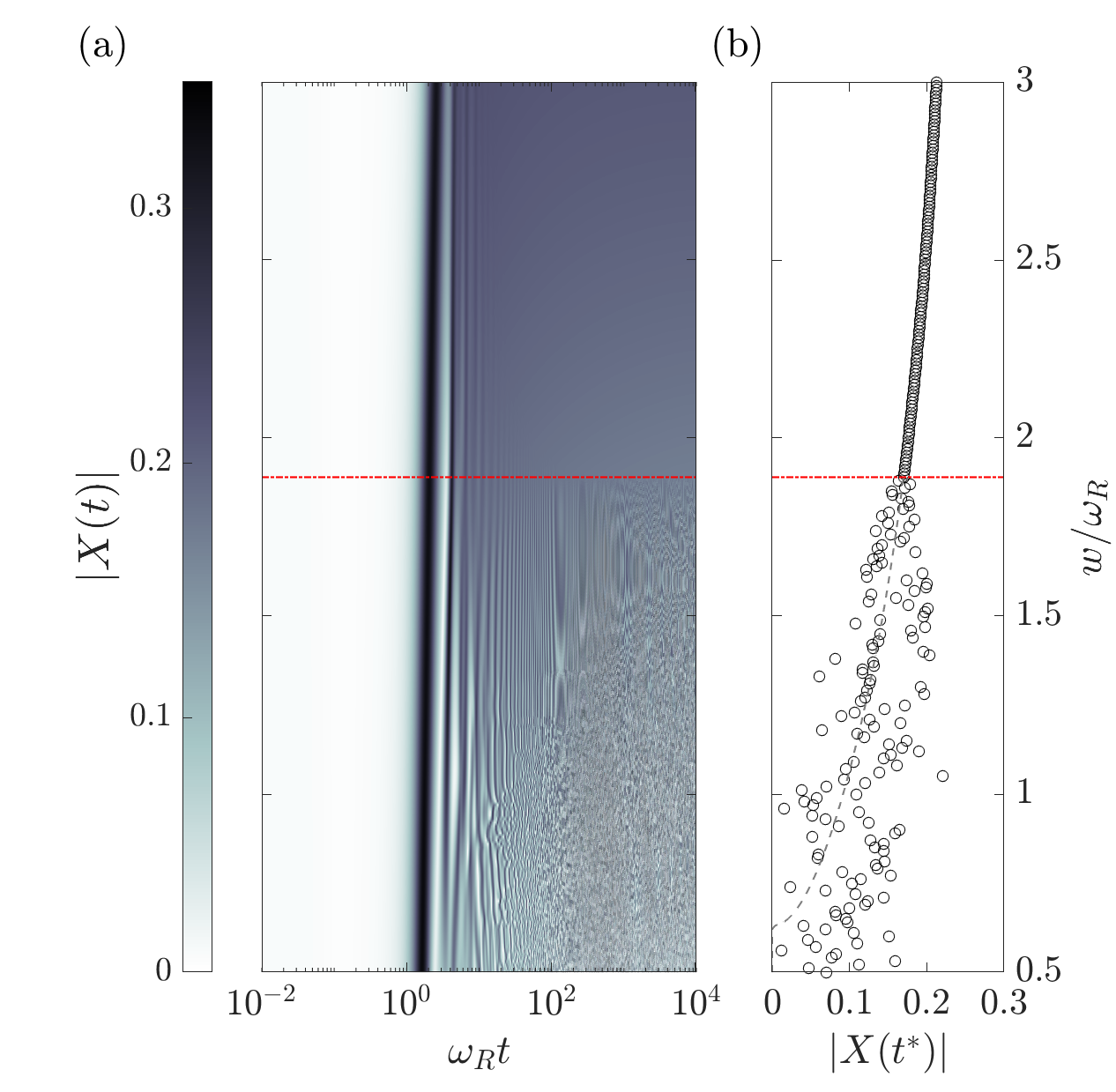}\\
	\caption{(a) Contour plot of $|X(t)|$ as a function of the time $t$ (in units of $\omega_R^{-1}$) and of $w$ (in units of $\omega_R$) for $\Lambda=15\omega_R$ and $\Delta=\kappa/2$. The dynamics of $|X(t)|$ is calculated by numerically integrating Eq.~\eqref{MEq:MF} with the initial state $\hat{\rho}_0$, Eq.~\eqref{Eq:BEC}, over a grid of momenta with a cut-off at $p_{\mathrm{max}}=15\hbar k$.  (b) The asymptotic value of $|X(t)|$ is displayed as a function of $w$ (in units of $\omega_R$). The dashed line is the solution of Eq. \eqref{steadystate}, the circles correspond to $X(t^*)$, as obtained by integrating Eq.~\eqref{MEq:MF} until $t^*=4\times 10^4\omega_R^{-1}$. The red dashed-dotted line in both subplots indicates the critical value $w_c$ at which the steady state solution of Eq.~\eqref{steadystate} predicts a transition from the chaotic to the coherent phase.\label{dynoscillation:X}}
\end{figure}

Figure~\ref{dynoscillation:X}(a) displays the contour plot of  $|X(t)|$  as a function of time and of $w$ for a constant value $\Lambda=15\omega_R>\Lambda_c$. The dynamics is initially characterized by the exponential buildup of $|X(t)|$, whose timescale weakly depends on $w$.  After this timescale the dynamics strongly depends on $w$. In particular, we identify a threshold value $w_c$ below which $|X(t)|$ performs large amplitude oscillations. In contrast, above $w_c$, $|X(t)|$ converges rapidly to its stationary value. The vertical line indicates the threshold value we extract from the solution of Eq.~\eqref{steadystate}. 
The features of this transition are also visible in the comparison of $|X(t^*)|$ evaluated at $t^*=10^4\omega_R^{-1}$ (circles) and the stationary values $X_{\rm st}$ found using Eq.~\eqref{steadystate} (dashed line) in Fig.~\ref{dynoscillation:X}(b). While for $w>w_c$, both results are in good agreement, for $w<w_c$, we observe discrepancies: $|X(t^*)|$ oscillates about its stationary value with an amplitude of the order of $|X|_{\rm st}$. 

\begin{figure}[h!]
    \center \includegraphics[width=1\linewidth]{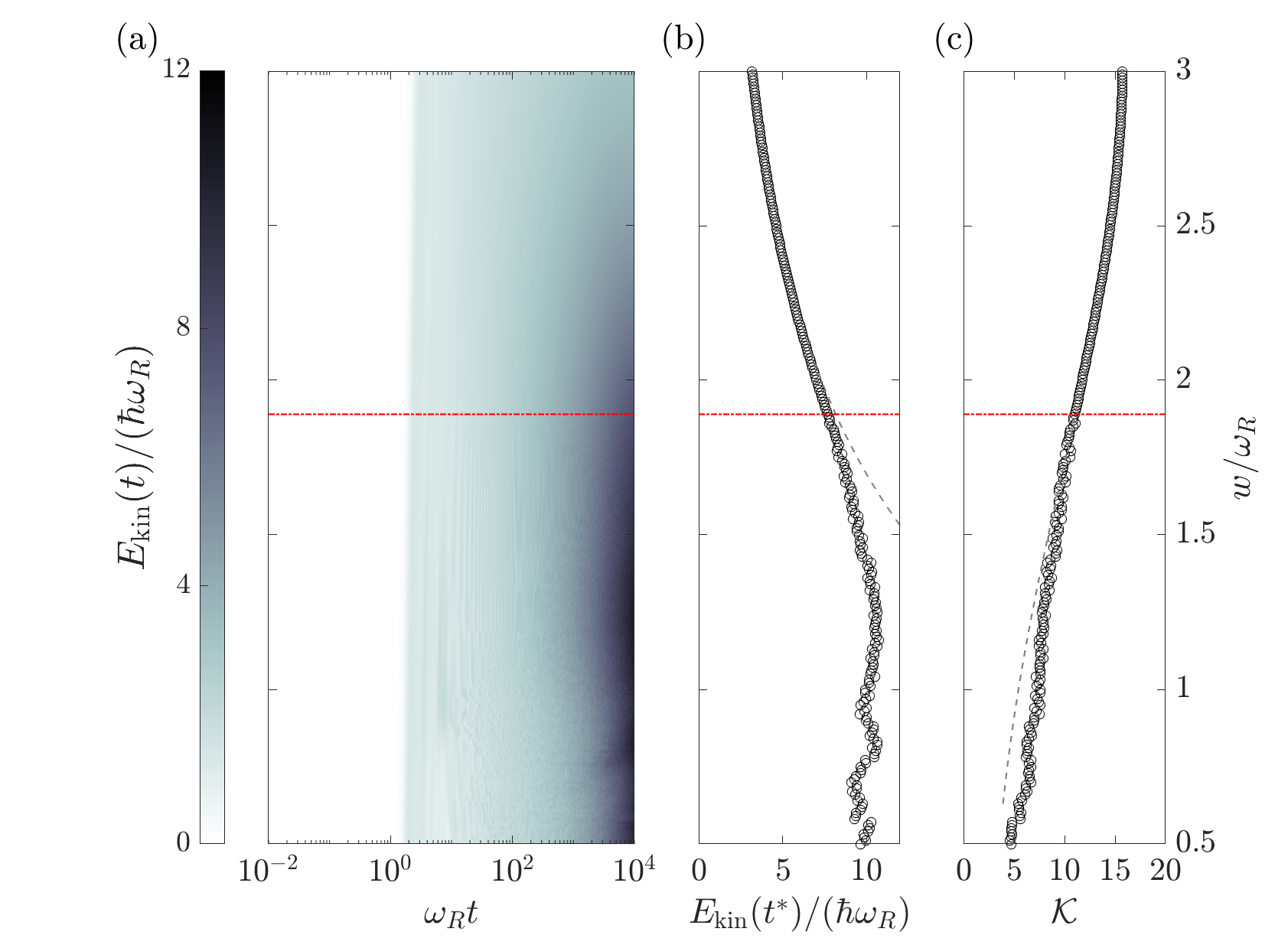}\\
	\caption{Subplots (a) and (b) are the same as in Fig.~\ref{dynoscillation:X} but for the kinetic energy $E_{\rm kin}=~\langle \hat{p}^2\rangle/(2m)$ (in units of the recoil energy $\hbar\omega_R$). Subplot (c) shows the kurtosis $\mathcal{K}(t)$ (Eq.~\eqref{Kurtosis}) at $t=t^*$ (circles) for different values of $w$ in units of $\omega_R$. The dashed line displays the corresponding value of the kurtosis, Eq.~\eqref{steadystate}.\label{dynoscillation:Kin}}
\end{figure}

The behavior of the kinetic energy is generally similar, as shown in Fig.~\ref{dynoscillation:Kin}(a), even though in the chaotic phase the oscillations are less pronounced than for $|X(t)|$. Subplot~(b) displays the asymptotic value: the kinetic energy is visibly higher in the chaotic phase. As for $|X(t)|$, for $w>w_c$, the mean value $E_{\rm kin,st}$ calculated by solving iteratively Eq.~\eqref{steadystate} agrees with the one found numerically. On the other hand, in the chaotic phase, $E_{\rm kin,st}>E_{\rm kin}(t^*)$ and the difference increases for smaller values of $w$. In subplot~(c) we show the kurtosis $\mathcal{K}(t)$ (Eq.~\eqref{Kurtosis}) evaluated at $t=t^*$ across the coherent-chaotic transition: the momentum distribution is non-Gaussian across the transition point~$w_c$ and for all considered values of $w$.

Numerical simulations show that for $w<w_c$ the density grating becomes unstable and the system jumps back and forth between a prevailing occupation of the set of states corresponding to an even grating, $\{\ket{e,\Psi_{2n}},\ket{g,\Psi_{2n+1}},n=~0,1,2,...\}$, and the set of states corresponding to an odd grating, $\{\ket{e,\Psi_{2n+1}},\ket{g,\Psi_{2n}},n=~0,1,2,...\}$. While the states within each set are coupled by coherent processes, the two sets are only coupled to each other by the incoherent pump: Thus, for $w<w_c$ the long-range optomechanical interactions tend to form a grating, which locks the phase of the field, while the incoherent pump induces quantum jumps between different gratings. An analysis of the entanglement is possible only from the coherent side, where the non-linear master equation has one stationary solution, and shows that internal and external degrees of freedom are entangled for $w>w_c$. 

\begin{figure}[h!]
	\center \includegraphics[width=0.9\linewidth]{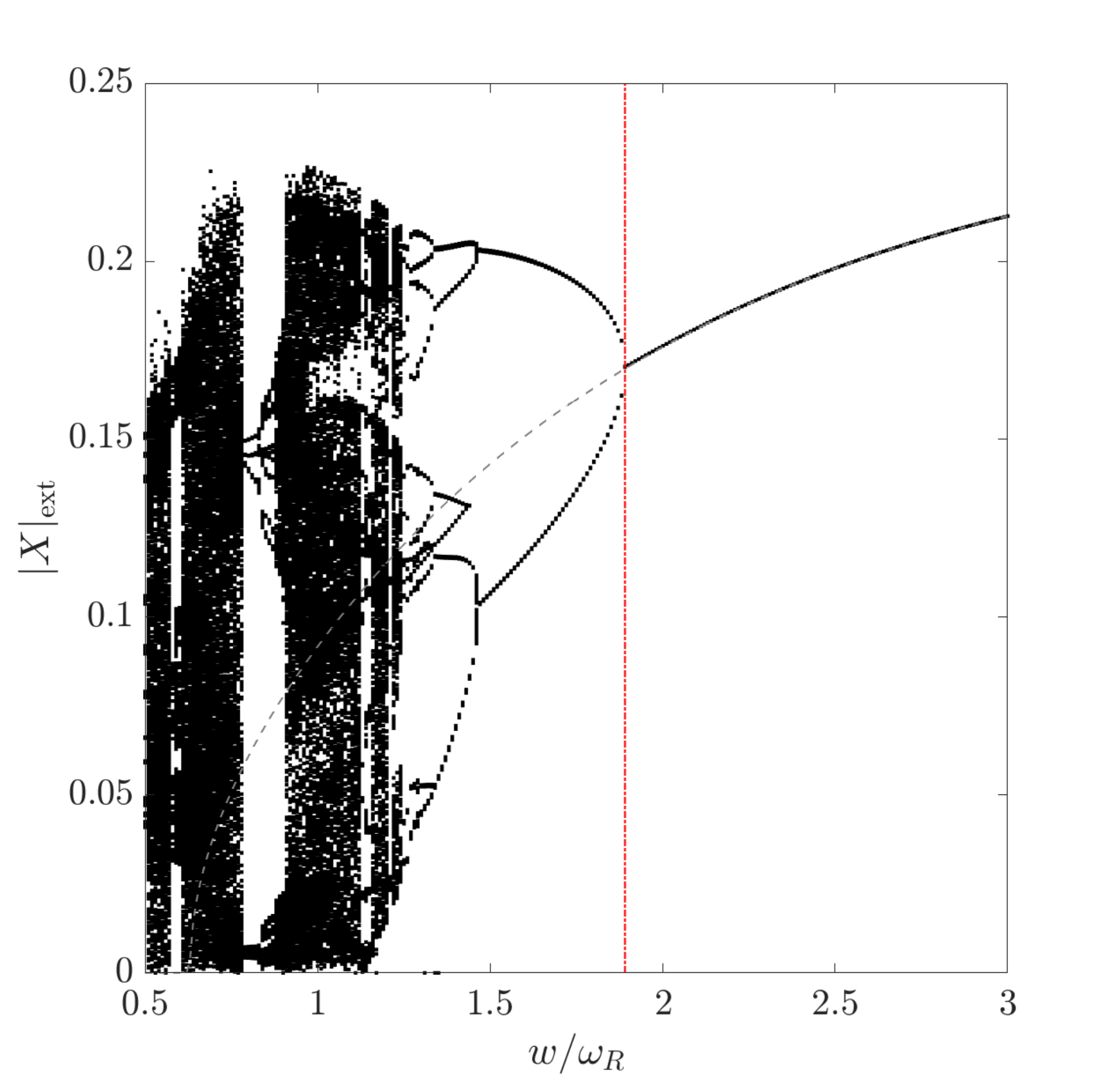}\\
	\caption{The value of the elements in the set $\Upsilon$ (Eq.~\ref{Upsilon}) for different values of $w$ (in units of $\omega_R$) evaluated in the interval $[t_0,t_1]$ with $t_0=9\times 10^3\omega_R^{-1}$ and $t_1=10^4\omega_R^{-1}$ (For $w>w_c$ we plot $|X(t^*)|$ for $t=10^4\omega_R^{-1}$). The red, dashed-dotted (gray dashed) line corresponds to the threshold value $w_c$ ($|X|_{\mathrm{st}}$) found by solving Eq.~\eqref{steadystate}. The other parameters are $\Lambda=15\omega_R$ and $\Delta =\kappa/2$. \label{fractal}}
\end{figure}

We now analyze whether the state of the system, which we have denoted as chaotic, indeed exhibits true signatures of chaos. We here note that in Ref.~\cite{Jaeger:2019} we performed a similar analysis of the spectrum as the one presented in subsection~\ref{relaxationthermalstate}. In Ref.~\cite{Jaeger:2019} we also characterized this transition by means of the Lyapunov exponent. In the following we present a further characterization. 
For this purpose, for a fixed value of $w$, we determine  the extrema of $|X(t)|$ in an appropriate  time interval $[t_0,t_1]$. We then define the set
\begin{align}
\Upsilon=\big\{|X|_{\mathrm{ext}}=|X(t)|\text{ is local extremum for }t\in[t_0,t_1]\big\}.\label{Upsilon}
\end{align}
The time interval is chosen such that the transient relaxation stage of $|X(t)|$ does not play a role. For every value of $w$ we plot all the points in $\Upsilon$. The results are displayed in Fig.~\ref{fractal} . 
For values $w>w_c$ we have checked that the set $\Upsilon$ is empty. Hence, here we have plotted $|X(t^*)|$, that is to a good approximation the stationary value. At $w\approx w_c$ we observe a bifurcation. By decreasing  $w$ we notice further bifurcations until the finite step size in $w$ limits the resolution. This might suggest that period doubling is leading the system from a single stationary state to chaotic dynamics. Another remarkable feature is the gap that occurs for $0.8 \lesssim w\lesssim 0.9$, where the set $\Upsilon$ shows only a few discrete points. Such features are known in chaos theory and can also for example be observed in the logistic map \cite{Strogatz}. 

\subsection{Spatio-temporal pattern}\label{spatiotemporal}

The phases resulting from the optomechanical coupling between internal and external degrees of freedom can be understood in terms of the stability of density gratings, which diffract the emitted light. The coherent regime, in particular, suggests a sort of phase locking between motion and spin, which we elucidate in this section. For this purpose we analyse the dynamics in time and position, and denote by $\{|x\rangle\}$ the eigenstates of the position operator $\hat x$. Since the Hamiltonian is periodic with period $\lambda=2\pi/k$ and we take periodic boundary conditions, it is sufficient to analyze the dynamics in the spatial interval~$(-\lambda/2,\lambda/2]$. 

Using the completeness relation we now write the order parameter as
\begin{align}
X(t)&=\frac{1}{\lambda}\int_{-\lambda/2}^{\lambda/2} dx\,{\rm Tr}\{\hat{\sigma}\cos(k\hat{x})\hat{\rho}_1(t)|x\rangle\langle x|\}\nonumber\\
&=\frac{1}{\lambda}\int_{-\lambda/2}^{\lambda/2} dx \,S(x,t)\,. \label{XfromS}
\end{align}
In the second line we have introduced the amplitude 
\begin{align}
S(x,t)&={\rm Tr}\{\hat{\sigma}\cos(k\hat{x})\hat{\rho}_1(t)|x\rangle\langle x|\}\,,\label{Spinxt}\\
&=|S(x,t)|e^{{\rm i}\phi(x,t)}\,.\label{Spinxt2}\,
\end{align}      
where $\phi(x,t)$ is its phase and $|S(x,t)|$ its modulus. In absence of optomechanical coupling this quantity is independent of position and $S$ reduces to the order parameter of synchronization \cite{Tucker:2018}. The mechanical effects of light induce the spatial dependence of the collective dipole and the superradiant phase can therefore be understood as a phase-locking between the dynamics of the internal and external degrees of freedom. In the coherent phase, where the system reaches stationary superradiance, this phase-locking shall be stationary. We analyze this behavior by 
studying the relative phase with respect to the phase at position $x=0$:
\begin{align}
\Delta\phi(x,t)=\phi(x,t)-\phi(0,t)\,.\label{Deltaphi}
\end{align}
This definition implies that $\Delta\phi(0,t)=0$ at every instant of time and for any initial condition. For $|S(x,t)|\neq 0$ for some position $x$, the necessary condition for which $|X(t)|$ reaches the maximum value at a given instant of time, is $\Delta\phi(x,t)=0$. 

\begin{figure}[h!]
	\center \includegraphics[width=0.9\linewidth]{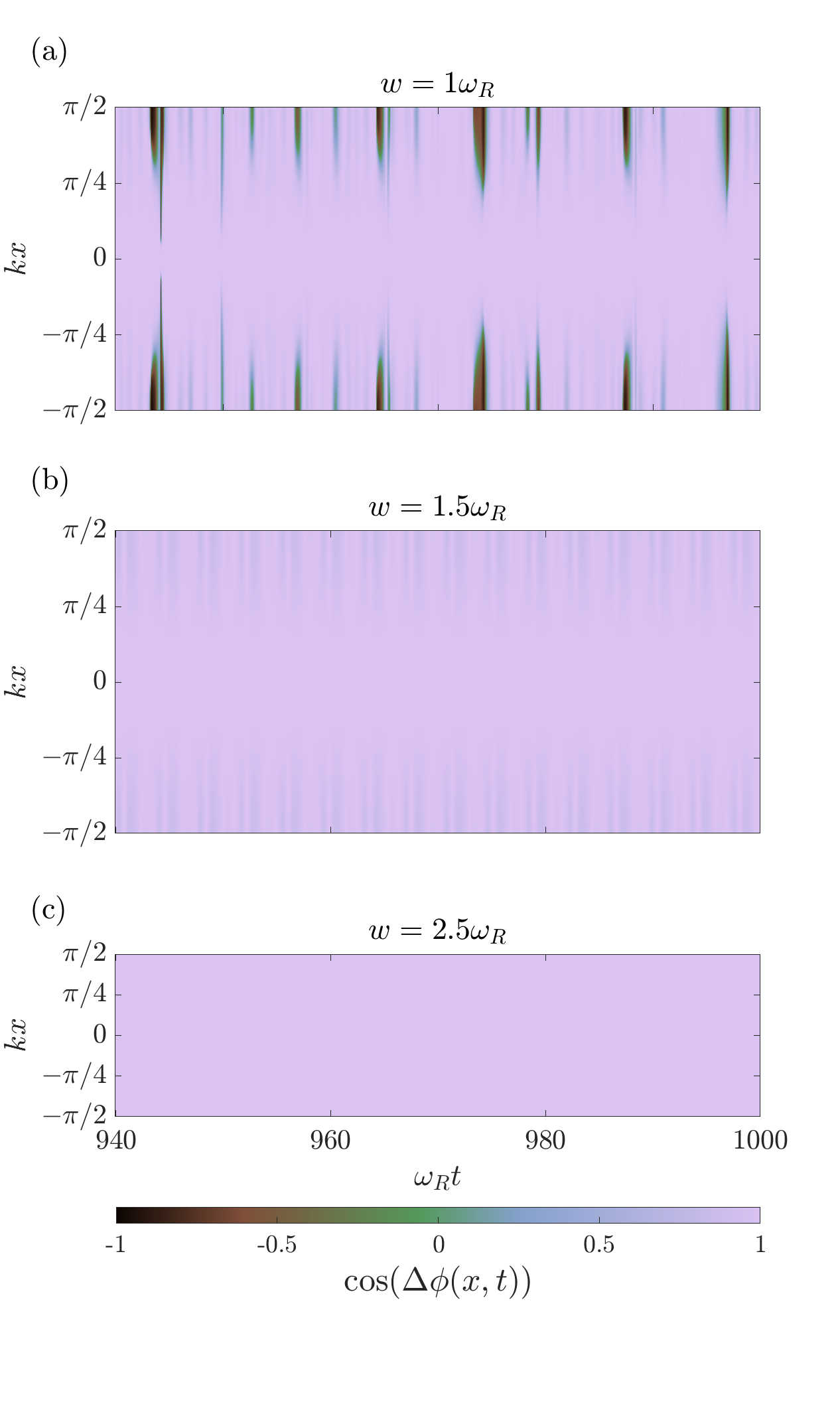}\\
	\caption{Dynamics of $\cos\Delta\phi(x,t)$, Eq.~\eqref{Deltaphi}, as a function of $x$ (in units $1/k$) for (a) $w=\omega_R$, (b) $w=1.5\omega_R$, and (c) $w=2.5\omega_R$. The other parameters are as in Fig.~\ref{dynoscillation:X} \label{phasedifference}}
\end{figure}

\begin{figure}[h!]
	\center \includegraphics[width=0.9\linewidth]{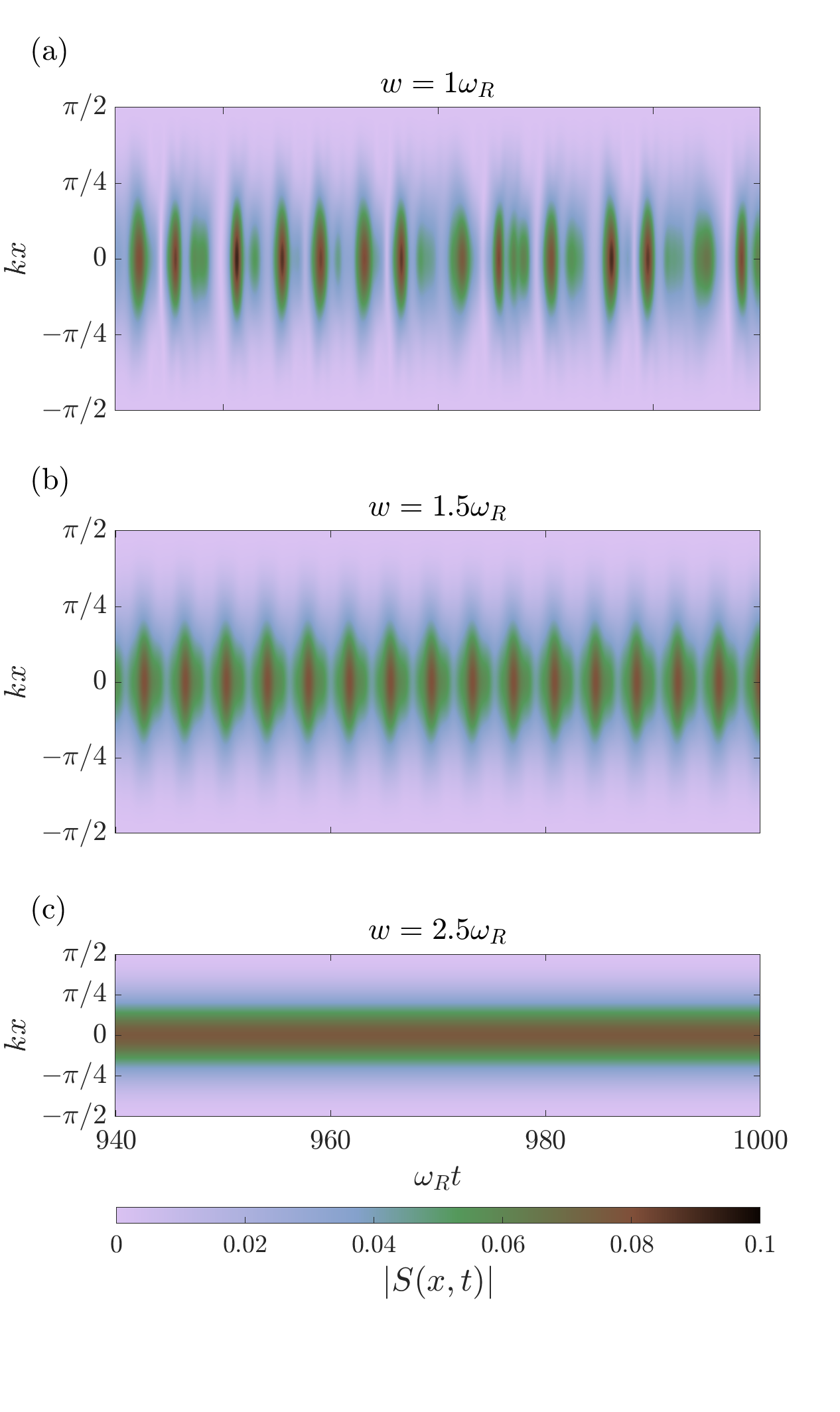}\\
	\caption{Same as in Fig.~\ref{phasedifference} but for the modulus $|S(x,t)|$. \label{Fig:modulus}}
\end{figure}

Figure~\ref{phasedifference} displays $\cos(\Delta\phi(x,t))$ for different values of $w$ across the chaotic-coherent transition. In the coherent phase (c), $\Delta \phi(x,t)$ has a constant value in space, indicating that motion and spin are stably phase locked. Close to the threshold value  $w_c$, as shown in subplot~(b), the phase starts to exhibit small amplitude oscillations in time and space about $x\sim \pm\lambda/2$. In subplot~(a), where $w<w_c$, one observes that the amplitude of these oscillations is maximum and the atoms at the positions $x\sim \pm\lambda/2$ become out of phase with respect to the atoms at $x=0$. The corresponding modulus of the collective spin is shown in Fig.~\ref{Fig:modulus}. These plots show that in the coherent phase (c), the atoms are stably trapped about $x=0$. At the transition point (b), the distribution starts to oscillate between being localized and centered, and spreading out over the whole interval. In subplot~(a) this behavior is enhanced and becomes irregular in time.  

These dynamics illustrate how spatial patterns accompany superradiant emission of light. In the chaotic phase the pattern is unstable, the appearance of the black stripes in Fig. \ref{phasedifference}(a) is accompanied by a redistribution of the atoms in a new pattern, which leads to a new configuration of phase locking. 

\begin{figure}[h!]
	\center \includegraphics[width=0.9\linewidth]{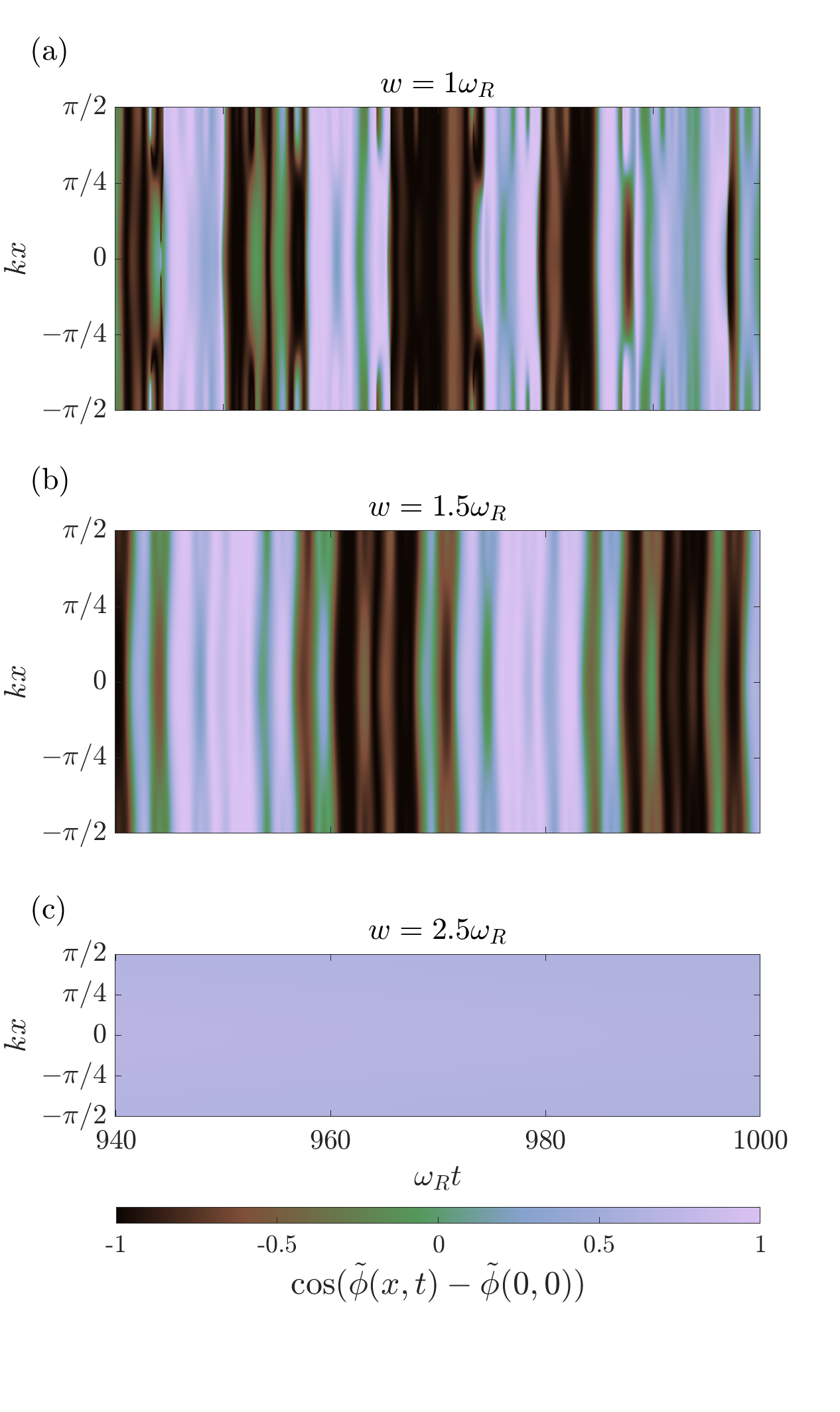}\\
	\caption{Same as in Fig.~\ref{phasedifference} but for $\cos(\tilde{\phi}(x,t)-\tilde{\phi}(0,0))$ and where $\tilde{\phi}$ is defined in Eq.~\eqref{tildephi}.\label{Fig:phi0}}
\end{figure}

This behavior is also visible in Fig.~\ref{Fig:phi0} where we show the dynamics of 
\begin{align}
\tilde{\phi}(x,t)=\phi(x,t)-\omega_wt\label{tildephi}
\end{align}
in the frame rotating with $\omega_{w}$ (see Eq.~\eqref{omegaw}) and with respect to the initial value $\tilde{\phi}(0,0)=\phi(0,0)$. As visible in Fig.~\ref{Fig:phi0}(c), in the coherent phase, $\tilde{\phi}(x,t)$ is spatially constant and almost stationary in time. This is in agreement with Eq.~\eqref{dphidt} that predicts that in the coherent phase there will be a stationary order parameter $X$ in the frame oscillating with frequency $\omega_w$.
In the chaotic phase, as is visible in Fig.~\ref{Fig:phi0}(a) and Fig.~\ref{Fig:phi0}(b), $\tilde{\phi}(x,t)$ shows several jumps that reflect the redistribution of the atoms and the corresponding temporal locking to a new phase.

\subsection{Experimental parameters}

The phase diagram we predict can be observed by tuning the superradiant linewidth $\Lambda$ and the pump rate $w$  across values of the order of the recoil frequency $\omega_R$. The phases are signaled by the first-order correlation function of the emitted light. These dynamics can be realized when the resonator linewidth $\kappa$ exceeds $\omega_R$ by several orders of magnitude and when other incoherent processes can be discarded over the timescales where the dynamical phase transition occurs. Specifically, the spontaneous decay of the dipolar transition and the particle-particle collision rate should be orders of magnitude smaller than the recoil frequency, which can be realized using a Raman transition between metastable hyperfine states and low densities, as for instance is demonstrated in Refs.~\cite{Bohnet:2012,Kroeze:2018,Landini:2018}. In order to provide some numbers, in the experiment of Ref.~\cite{Bohnet:2012} the cavity linewidth is $\kappa\sim 11$ MHz, and the recoil frequency is approximately $\omega_R\sim 4$ kHz. For a single-atom cooperativity  $C\sim10^{-2}$, a tunable effective single-particle linewidth $\gamma\sim 0.2-10$ Hz, and a particle number $N\sim10^6$, one obtains for the collective linewidth $\Lambda\sim NC\gamma=2-100$ kHz, which is in the parameter regime for observing the dynamics we predict. 

A further note regards the effect of $s$-wave collisions, which we have discarded in our treatment. For a BEC of Rb$^{87}$ atoms with a density $n\sim 10^{14}$cm$^{-3}$ \cite{Ketterle:2006} and a scattering length $a\sim 100 a_0$ \cite{Scattering:Length}, with $a_0$ the Bohr radius, we expect a mean-field energy shift corresponding to $gn/h=2\hbar a_s n/m\sim10$ Hz, which can be neglected for the timescales we consider. Note that in our simulations we also neglect the contribution of the atomic linewidth $\gamma$ of the excited state.

\section{Conclusions}
\label{Sec:conclusions}

We have analyzed the stability of stationary superradiance in a standing-wave optical resonator when the optomechanical coupling to the atomic motion is relevant to the dynamics. For this purpose, we derived a mean-field model and studied its predictions for a specific class of initial states. Our model was formulated for a situation where the atoms are all prepared in the excited state (the medium exhibits perfect population inversion) and the motion is in a thermal state. A stability analysis allowed us to show that the parameter regime, where the system can undergo a superradiant emission, depends on the temperature. Moreover, we identified a critical initial temperature above which superradiance is suppressed by the atomic motion. 

We then studied the case where the atomic motion can be described by a semiclassical model and when instead the atoms form an ideal Bose-Einstein condensed gas at $T=0$. Our mean-field treatment allowed us to identify three different phases in the regime where one expects stationary superradiance in the absence of motional effects. We indeed recovered stationary superradiance when the relevant rates (the superradiant emission rate and the incoherent pump rate) exceed the recoil frequency. On the contrary, when they are comparable to the recoil frequency, we identified threshold values below which the system either dephases and stops emitting (the ``incoherent'' phase), or it emits light which exhibits chaotic features. This chaotic behavior is found also in the semiclassical regime.
We provided illustrative pictures of the dynamics, showing that the onset of stationary superradiance is accompanied by the formation of a stable spatial pattern which diffracts the emitted light, while in the chaotic phase there is an interplay between the coherent emission, which establishes a pattern locked to the phase of the field, and the incoherent pump, which destroys the phase coherence. The dynamics shares analogy with superradiant emission in a Bose-Einstein condensate \cite{Inouye:1999,Schneble:2003,Baumann:2010,Safei:2013}, where superradiant gain can be understood as the diffraction of photons from the density grating of the recoiling atoms, that acts as an amplifying medium. In the case discussed here there is no phase coherence between the incident and the emitted light, the pump is incoherent and the gratings are spontaneously formed as a result of the interplay between the noise of the pump, the long-range interactions mediated by the cavity photons, and the dissipation due to cavity losses. 

We emphasize that in the asymptotic limit of the dynamics the atoms do not form a Bose-Einstein condensate (BEC); in the incoherent phase the atomic state is a statistical mixture, in the chaotic phase the atoms temporally jump between different gratings, and in the coherent phase they form a grating. None of these states can be correctly classified as a BEC. Moreover, the transition we observe is within the class of metastable states that are reached by starting from a BEC with all atoms in the excited state. We remark that the phases would be generally different for different classes of initial states, as is typical in the physics of long-range (non-additive) interacting systems \cite{Campa:2009}. The chaotic phase, in particular, characterizes the asymptotic phase of an incoherent dynamics, it emerges from the interplay between quantum fluctuations, noise, and all-to-all, global interactions mediated by the cavity field, and is thus qualitatively different from chaos reported in the quantum dynamics of Hamiltonian all-to-all, global-range interacting systems \cite{Emary:2003,Lerose:2018}. It is intriguing whether it presents features which can be understood in terms of many-body quantum chaos \cite{Carr:2019}.

\acknowledgements
We thank John Cooper for stimulating discussions and helpful comments.
This work has been supported by the German Research Foundation (DACH ``Quantum crystals of matter and light'' and the priority program No. 1929 GiRyd), by the European
Commission (ITN ColOpt) and by the German Ministry of Education and Research (BMBF) via the QuantERA project NAQUAS. Project NAQUAS has
received funding from the QuantERA ERA-NET Cofund in Quantum Technologies implemented within the European Union's Horizon 2020 program. We would like to acknowledge support from
the NSF PFC Grant No. PHY 1734006, NSF Grant No. PHY
1806827 and the DARPA Extreme Sensing program.

\appendix

\section{Derivation of the master equation} \label{App:A}   
The dynamics of the density matrix $\hat{\rho}$ is governed by the Lindbladian
\begin{align*}
\frac{\partial \hat{\rho}}{\partial t}=\mathcal{L}\hat{\rho}.
\end{align*}
The Lindbladian $\mathcal{L}$ can be cast into three terms
\begin{align*}
\mathcal{L}=\mathcal{L}_F+\mathcal{L}_A+\mathcal{L}_1,
\end{align*}
where the three terms take the forms
\begin{align*}
\mathcal{L}_F\hat{\rho}=&\frac{1}{{\rm i}\hbar}\left[\hbar\left(\Delta-{\rm i}\frac{\kappa}{2}\right)\hat{a}^{\dag}\hat{a}\right]\hat{\rho}+\mathrm{H.c.}+\kappa\hat{a}\hat{\rho}\hat{a}^{\dag},\\
\mathcal{L}_A\hat{\rho}=&\frac{1}{{\rm i}\hbar}\left[\frac{\hat{p}_j^2}{2m}-{\rm i}\hbar\frac{w}{2}\sum_j\hat{\sigma}_j\hat{\sigma}_j^{\dag}\right]\hat{\rho}+\mathrm{H.c.}+w\sum_j\hat{\sigma}_j^{\dag}\hat{\rho}\hat{\sigma}_j,\\
\mathcal{L}_1\hat{\rho}=&\frac{1}{{\rm i}\hbar}\left[\sum_{j}\frac{\hbar g\cos(k\hat{x}_j)}{2}(\hat{a}^{\dag}\hat{\sigma}_j+\hat{\sigma}_j^{\dag}\hat{a})\right]\hat{\rho}+\mathrm{H.c.}.
\end{align*}
The Lindbladian $\mathcal{L}_F$ describes dynamics of the cavity degrees of freedom while $\mathcal{L}_A$ describes the dynamics of the external and internal degrees of freedom of the atoms. The last term $\mathcal{L}_1$ is a coupling between cavity and atomic degrees of freedom. The purpose of this appendix is to eliminate the cavity degrees of freedom under the assumption that the coupling of cavity and atomic degrees of freedom is weak in comparison to the cavity decay rate $\kappa$. Moreover we assume that the atomic degrees of freedom evolve very slowly such that also the Doppler shift $k\Delta p/m$ and the incoherent pump rate $w$ are much smaller than $\kappa$. 

In order to derive an effective Master equation for the atomic degrees of freedom we use the projector method\cite{Gardiner:QuantumNoise}. For this purpose, we define the projector 
$$\mathcal{P}\hat{\rho}=|0\rangle\langle 0|\hat{\rho}|0\rangle\langle 0|\equiv \hat{\rho}_{\rm red}\otimes |0\rangle\langle 0|\,,$$ with $\hat{\rho}_{\rm red}=\langle 0|\hat{\rho}|0\rangle$, and the projector $\mathcal{Q}=1-\mathcal{P}$ to its orthogonal space. We define further the matrices $\hat{v}=\mathcal{P}\hat{\rho}$ and $\hat{w}=\mathcal{Q}\hat{\rho}$. Applying now these projectors onto the master equation we obtain two coupled equations for $\hat{v}$ and $\hat{w}$:
\begin{align}
\frac{\partial \hat{v}}{\partial t}=&\mathcal{P}\mathcal{L}_F\hat{w}+\mathcal{P}\mathcal{L}_A\hat{v}+\mathcal{P}\mathcal{L}_1\hat{w},\label{dvdt}\\
\frac{\partial \hat{w}}{\partial t}=&\mathcal{Q}\mathcal{L}_F\hat{w}+\mathcal{Q}\mathcal{L}_A\hat{w}+\mathcal{Q}\mathcal{L}_1(\hat{v}+\hat{w}).\label{dwdt}
\end{align}
We used here that $\mathcal{L}_F\hat{v}=0$ and that $\mathcal{L}_A$ commutes with the corresponding projectors.
All frequencies are taken in reference to $\kappa$ and $\Delta$, and thus we use the following hierarchy of orders of magnitudes. For the cavity degrees of freedom
\begin{align*}
\mathcal{L}_{F}\hat{\rho}\sim&\mathrm{O}(1).
\end{align*}
The approximation that the timescales between atomic motion and cavity motion can be separated can be written as 
\begin{align*}
\mathcal{L}_{A}\hat{\rho}\sim&\mathrm{O}(\epsilon),
\end{align*}
where $\epsilon\sim k\Delta p/m/\kappa,w/\kappa$ should be very small. If we furthermore assume that the coupling between cavity and atomic degrees of freedom is weak we may write
\begin{align*}
\mathcal{L}_{1}\hat{\rho}\sim&\mathrm{O}(\epsilon).
\end{align*}
Here, we have used that $\epsilon\sim g\sqrt{\bar n}/\kappa$ where $\bar{n}$ denotes the mean intracavity photon number. Hence a large photon number would violate this assumption. Our goal is to derive a master equation that is correct to order $\mathcal{O}(\epsilon^2)$.

We now formally integrate $\hat{w}$ and obtain a result that it is of first order in $\epsilon$
\begin{align*}
\hat{w}(t)=\int_{t_0}^{t}d\tau\,e^{\mathcal{Q}(\mathcal{L}_F+\mathcal{L}_A+\mathcal{L}_1)(t-\tau)}\mathcal{Q}\mathcal{L}_1\hat{v}(\tau)\sim\mathrm{O}(\epsilon).
\end{align*}
This equation is now used in Eq.~\eqref{dvdt} where we get the following result
\begin{align}
\frac{\partial \hat{v}}{\partial t}=&\mathcal{P}\mathcal{L}_A\hat{v}(t)+\mathcal{P}\mathcal{L}_1\int_{0}^{\Delta t}d\tau\,e^{\mathcal{Q}(\mathcal{L}_F+\mathcal{L}_A+\mathcal{L}_1)\tau}\mathcal{Q}\mathcal{L}_1\hat{v}(t-\tau)\nonumber\\
&+\mathcal{P}\mathcal{L}_F\int_{0}^{\Delta t}d\tau\,e^{\mathcal{Q}(\mathcal{L}_F+\mathcal{L}_A+\mathcal{L}_1)\tau}\mathcal{Q}\mathcal{L}_1\hat{v}(t-\tau)\label{vt}.
\end{align}
Since the second term in the first line of Eq.~\eqref{vt} is already second order in $\epsilon$ we can drop the terms $\mathcal{L}_A$ and $\mathcal{L}_1$ in the exponential. In the second line of Eq.~\eqref{vt} we expand the exponential up to first order in $\epsilon$ to keep consistently all terms up to second order in $\epsilon$ in Eq.~\eqref{vt}. We use the expansion of the propagator
\begin{align}
e^{\mathcal{Q}(\mathcal{L}_F+\mathcal{L}_A+\mathcal{L}_1)\tau}\approx \Phi_0+\Phi_1,\label{expansion}
\end{align} 
where 
\begin{align}
\Phi_0=e^{\mathcal{Q}\mathcal{L}_F\tau}
\end{align} 
describes the zeroth order in $\epsilon$ and 
\begin{align}
\Phi_1=&\frac{d}{d\varepsilon}\left. e^{\mathcal{Q}(\mathcal{L}_F+\varepsilon[\mathcal{L}_A+\mathcal{L}_1])\tau}\right|_{\varepsilon=0}\\
=&\int_0^{\tau}d\tau'e^{\mathcal{Q}\mathcal{L}_F\tau'}[\mathcal{L}_A+\mathcal{L}_1]e^{\mathcal{Q}\mathcal{L}_F(\tau-\tau')}
\end{align}
is the first order in $\epsilon$. Using these expressions in Eq.~\eqref{vt} we obtain 
\begin{align}
&\frac{\partial \hat{v}}{\partial t}=\mathcal{P}\mathcal{L}_A\hat{v}(t)+\mathcal{P}\mathcal{L}_1\int_{0}^{\Delta t}d\tau\,e^{\mathcal{Q}\mathcal{L}_F\tau}\mathcal{Q}\mathcal{L}_1\hat{v}(t-\tau)\nonumber\\
&+\mathcal{P}\mathcal{L}_F\int_{0}^{\Delta t}d\tau\int_0^{\tau}d\tau'\,e^{\mathcal{Q}\mathcal{L}_F\tau'}\mathcal{L}_1e^{\mathcal{Q}\mathcal{L}_F(\tau-\tau')}\mathcal{Q}\mathcal{L}_1\hat{v}(t-\tau)\label{v},
\end{align}
where we have already used the fact that the contribution of $\Phi_0$ and $\mathcal{L}_A$ in $\Phi_1$ vanish when applying the expansion (Eq.~\eqref{expansion}) and inserting it in Eq.~\eqref{vt}. Because of the timescale separation we may replace $\hat{v}(t-\tau)$ with $\hat{v}(t)$ and may exchange the integration time $\Delta t$ by infinity. Using $\hat{v}=\hat{\rho}_{\mathrm{red}}\otimes|0\rangle\langle 0|$ and tracing out the cavity field we finally get the result shown in Eq. \eqref{Mastereq1}.

\section{Time evolution of $\hat{g}_2$}\label{App:B}

The equation of motion for $\hat{g}_2$ takes the form 
\begin{align}
\frac{\partial \hat{g}_2}{\partial t}=\frac{\partial \hat{\rho}_2}{\partial t}-\frac{\partial \hat{\rho}_1}{\partial t}\otimes \hat{\rho}_1-\hat{\rho}_1\otimes \frac{\partial \hat{\rho}_1}{\partial t}.
\end{align}
Using now Eq.~\eqref{Masterequl} for $\ell=1$ and $\ell=2$ and Eq.~\eqref{decompositionrho2} we decompose the equation into the sum of two terms:
\begin{align}
\frac{\partial \hat{g}_2}{\partial t}=&\hat{A}_1+\hat{A}_2\,.\label{g2dynamics}
\end{align}

We first report $\hat{A}_1$. We embed a single particle operator $\hat{O}_1$ into the two-particle space by defining $\hat{O}_1^{(1)}=~\hat{O}_1\otimes \hat{1}$ and $\hat{O}_1^{(2)}=\hat{1}\otimes \hat{O}_1$ where $\hat{1}$ is the identity operator on the other particle space, respectively.
With this definition $\hat{A}_1$ can be written as
\begin{align*}
\hat{A}_1=&\frac{1}{{\rm i}\hbar}\left[\sum_{l=1}^2\frac{\hat{p}_l^2}{2m},\hat{g}_2\right]-\frac{1}{{\rm i}\hbar}\left[\sum_{j=1}^2\frac{\hbar \Lambda}{2\sin\chi N}\hat{J}_2^{\dag}\hat{J}_2,\hat{g}_2\right]\\
&+\frac{1}{{\rm i}\hbar}\left[-\frac{\hbar\Lambda}{2\sin\chi N}((\hat{J}_1^{(1)})^{\dag}\hat{J}_1^{(2)}+(\hat{J}_1^{(2)})^{\dag}\hat{J}_1^{(1)}),\hat{\rho}_1\otimes\hat{\rho}_1\right]\\
&-\frac{\Lambda}{2N}\left([\hat{J}_1^{(1)})^{\dag},\hat{J}_1^{(2)}\hat{\rho}_1\otimes\hat{\rho}_1]+[\hat{\rho}_1\otimes\hat{\rho}_1 (\hat{J}_1^{(1)})^{\dag},\hat{J}_1^{(2)}]\right)\\
&-\frac{\Lambda}{2N}\left([(\hat{J}_1^{(2)})^{\dag},\hat{J}_1^{(1)}\hat{\rho}_1\otimes\hat{\rho}_1]+[\hat{\rho}_1\otimes\hat{\rho}_1 (\hat{J}_1^{(2)})^{\dag},\hat{J}_1^{(1)}]\right)\\
&-\sum_{j=1}^2\frac{w}{2}\left(\hat{\sigma}_j\hat{\sigma}_j^{\dag}\hat{g}_2+\hat{g}_2\hat{\sigma}_j\hat{\sigma}_j^{\dag}-2\hat{\sigma}_j^{\dag}g_2\hat{\sigma}_j\right)\\
&-\sum_{j=1}^2\frac{\Lambda}{2N}\left(\hat{J}_2^{\dag}\hat{J}_2\hat{g}_2+\hat{g}_2\hat{J}_2^{\dag}\hat{J}_2-2\hat{J}_2\hat{g}_2\hat{J}_2^{\dag}\right).
\end{align*}
The explicit form of $\hat{A}_2$ is given by
\begin{align*}
\hat{A}_2=&{\rm i}\left(1-\frac{2}{N}\right)\frac{\Lambda}{2\sin\chi }\left({\rm e}^{-{\rm i}\chi}[\hat{J}_2,\hat{\mathcal{X}}_2^*[\hat{\rho}_3]] -\mathrm{H.c.}\right)\\
&-{\rm i}\left(1-\frac{1}{N}\right)\frac{\Lambda}{2\sin\chi }\left({\rm e}^{-{\rm i}\chi}[\hat{J}_1,\hat{\mathcal{X}}_1^*[\hat{\rho}_2]]\otimes\hat{\rho}_1 -\mathrm{H.c.}\right)\\
&-{\rm i}\left(1-\frac{1}{N}\right)\frac{\Lambda}{2\sin\chi }\left({\rm e}^{-{\rm i}\chi}\hat{\rho}_1\otimes[\hat{J}_1,\hat{\mathcal{X}}_1^*[\hat{\rho}_2]] -\mathrm{H.c.}\right).
\end{align*}
We now decompose $\hat \rho_3$ as
\begin{align}
\hat{\rho}_3=\hat{\rho}_1\otimes\hat{\rho}_1\otimes \hat{\rho}_1+\sum_{j,k,l=1}^3|\epsilon_{jkl}|\hat{h}_{j,kl}+\hat{g}_3.\label{decompositionrho3}
\end{align}
Here $\hat{h}_{j,kl}$ describes a correlation between the $k$th and $l$th Hilbert space, with no correlation to $j$, and $\epsilon_{jkl}$ is the Levi-Civita tensor. The matrix $\hat{g}_3$ describes three particle correlations. The quantity $h_{j,kl}$ together with an arbitrary single particle operator $\hat{A}$ acting on the $j$th Hilbert space and an arbitrary two particle operator $\hat{B}$ acting on the $k$ and $l$th Hilbert space fulfills the condition
\begin{align}
\mathrm{Tr}(\hat{A}\hat{B}\hat{h}_{j,kl})=\mathrm{Tr}(\hat{A}\hat{\rho}_1)\mathrm{Tr}(\hat{B}\hat{g}_2).\label{hrules}
\end{align}
Using the decomposition of $\hat{\rho}_3$ in Eq. \eqref{decompositionrho3} we get
$$\hat{A}_2=\frac{\Lambda}{2\sin\chi}\hat a_2\,,$$
with
\begin{align*}
\hat{a}_2=&{\rm i}\left(1-\frac{2}{N}\right)\left({\rm e}^{-{\rm i}\chi}[\hat{J}_2,X^*\hat{\rho}_1\otimes\hat{\rho}_1] -\mathrm{H.c.}\right)\\
&+{\rm i}\left(1-\frac{2}{N}\right)\sum_{j,k,l=1}^3|\epsilon_{jkl}|\left({\rm e}^{-{\rm i}\chi}[\hat{J}_2,\hat{\mathcal{X}}_2^*[\hat{h}_{j,kl}]] -\mathrm{H.c.}\right)\\
&+{\rm i}\left(1-\frac{2}{N}\right)\left({\rm e}^{-{\rm i}\chi}[\hat{J}_2,\hat{\mathcal{X}}_2^*[\hat{g}_3]] -\mathrm{H.c.}\right)\\
&-{\rm i}\left(1-\frac{1}{N}\right)\left({\rm e}^{-{\rm i}\chi}[\hat{J}_1^{(1)},X^*\hat{\rho}_1\otimes\hat{\rho}_1] -\mathrm{H.c.}\right)\\
&-{\rm i}\left(1-\frac{1}{N}\right)\left({\rm e}^{-{\rm i}\chi}[\hat{J}_1^{(2)},\hat{\rho}_1\otimes\hat{\rho}_1X^*] -\mathrm{H.c.}\right)\\
&-{\rm i}\left(1-\frac{1}{N}\right)\left({\rm e}^{-{\rm i}\chi}[\hat{J}_1^{(1)},\hat{\mathcal{X}}_1^*[\hat{g}_2]\otimes\rho_1] -\mathrm{H.c.}\right)\\
&-{\rm i}\left(1-\frac{1}{N}\right)\left({\rm e}^{-{\rm i}\chi}[\hat{J}_1^{(2)},\hat{\rho}_1\otimes\hat{\mathcal{X}}_1^*[\hat{g}_2]] -\mathrm{H.c.}\right).
\end{align*}
We now use Eqs. \eqref{hrules} to obtain
\begin{align*}
\hat{a}_2
=&-{\rm i}\frac{1}{N}[{\rm e}^{-{\rm i}\chi} X^* \hat{J}_2+{\rm e}^{{\rm i}\chi}\hat{J}_2^{\dag}X,\hat{\rho}_1\otimes\hat{\rho}_1] \\
&+{\rm i}(1-2/N)[{\rm e}^{-{\rm i}\chi} X_0^* \hat{J}_2+{\rm e}^{{\rm i}\chi}\hat{J}_2^{\dag}X,\hat{g}_2] \\
&+{\rm i}(1-1/N)\left({\rm e}^{-{\rm i}\chi}\hat{\mathcal{X}}_1^*[\hat{g}_2]\otimes [\hat{J}_1,\hat{\rho}_1] -\mathrm{H.c.}\right)\\
&+{\rm i}(1-1/N)\left({\rm e}^{-{\rm i}\chi}[\hat{J}_1,\hat{\rho}_1]\otimes\hat{\mathcal{X}}_1^*[\hat{g}_2] -\mathrm{H.c.}\right)\\
&-{\rm i}\frac{1}{N}\left({\rm e}^{-{\rm i}\chi}[\hat{J}_2,\hat{\mathcal{X}}_1^*[\hat{g}_2]\otimes\hat{\rho}_1] -\mathrm{H.c.}\right)\\
&-{\rm i}\frac{1}{N}\left({\rm e}^{-{\rm i}\chi}[\hat{J}_2,\hat{\rho}_1\otimes\hat{\mathcal{X}}_1^*[\hat{g}_2]] -\mathrm{H.c.}\right)\\
&+{\rm i}(1-2/N)\left({\rm e}^{-{\rm i}\chi}[\hat{J}_2,\hat{\mathcal{X}}_2^*[\hat{g}_3]] -\mathrm{H.c.}\right),\\
\end{align*} 
where we used the decomposition $\hat{J}_2=\hat{J}_1^{(1)}+\hat{J}_1^{(2)}$. If we now assume that $\hat{g}_2$ and $\hat{g}_3$ are at least of order $1/N$ we obtain Eq. \eqref{g2eq}.

\begin{widetext}
\section{Supplemental information to the stability analysis}\label{App:C}

The entries of the matrix in Eq.~\eqref{CMatrix} are
\begin{eqnarray}
C_{11}=&-{\rm i}\frac{\Lambda {\rm e}^{{\rm i}\chi}}{2\sin\chi}\mathrm{Tr}\left\{\hat{J}_1\left(s-\tilde{\mathcal{L}}_{\mathrm{mf}}\right)^{-1}[\hat{J}_1^{\dag},\hat{\tilde\rho}_X]\right\}&=-{\rm i}\frac{\Lambda {\rm e}^{{\rm i}\chi}}{2\sin\chi}\int_{0}^\infty\,dt\, e^{-st}\langle[\hat{J}_1(t),\hat{J}_1^{\dag}(0)]\rangle,\label{C11}\\
C_{12}=&-{\rm i}\frac{\Lambda{\rm e}^{-{\rm i}\chi}}{2\sin\chi}\mathrm{Tr}\left\{\hat{J}_1\left(s-\tilde{\mathcal{L}}_{\mathrm{mf}}\right)^{-1}[\hat{J}_1,\hat{\tilde\rho}_X]\right\}&=-{\rm i}\frac{\Lambda {\rm e}^{-{\rm i}\chi}}{2\sin\chi}\int_{0}^\infty\,dt\, e^{-st}\langle[\hat{J}_1(t),\hat{J}_1(0)]\rangle,\\
C_{21}=&-{\rm i}\frac{\Lambda{\rm e}^{{\rm i}\chi}}{2\sin\chi}\mathrm{Tr}\left\{\hat{J}_1^{\dag}\left(s-\tilde{\mathcal{L}}_{\mathrm{mf}}\right)^{-1}[\hat{J}_1^{\dag},\hat{\tilde\rho}_X]\right\}&=-{\rm i}\frac{\Lambda {\rm e}^{{\rm i}\chi}}{2\sin\chi}\int_{0}^\infty\,dt\, e^{-st}\langle[\hat{J}_1^{\dag}(t),\hat{J}_1^{\dag}(0)]\rangle,\\
C_{22}=&-{\rm i}\frac{\Lambda{\rm e}^{-{\rm i}\chi}}{2\sin\chi}\mathrm{Tr}\left\{\hat{J}_1^{\dag}\left(s-\tilde{\mathcal{L}}_{\mathrm{mf}}\right)^{-1}[\hat{J}_1,\hat{\tilde\rho}_X]\right\}&=-{\rm i}\frac{\Lambda {\rm e}^{-{\rm i}\chi}}{2\sin\chi}\int_{0}^\infty\,dt\, e^{-st}\langle[\hat{J}_1(t),\hat{J}_1^{\dag}(0)]\rangle\,.
\end{eqnarray}
The elements of vector ${\bf b}$ read:
\begin{eqnarray}
b_{1}=&\mathrm{Tr}\left\{\hat{J}_1\left(s-\tilde{\mathcal{L}}_{\mathrm{mf}}\right)^{-1}\delta\hat{\tilde\rho}(0)\right\}&=\int_{0}^\infty \,dt\,e^{-st}\mathrm{Tr}\left\{\hat{J}_1(t)\delta\hat{\tilde\rho}(0)\right\},\\
b_{2}=&\mathrm{Tr}\left\{\hat{J}_1^{\dag}\left(s-\tilde{\mathcal{L}}_{\mathrm{mf}}\right)^{-1}\delta\hat{\tilde\rho}(0)\right\}&=\int_{0}^\infty \,dt\,e^{-st}\mathrm{Tr}\left\{\hat{J}_1^{\dag}(t)\delta\hat{\tilde\rho}(0)\right\}.
\end{eqnarray}
Here, we have used that $\tilde{\mathcal{L}}_{\mathrm{mf}}=\tilde{\mathcal{L}}_{\mathrm{mf}}[\hat{\tilde\rho}_X]$ is time independent and the Heisenberg picture to define
\begin{align*}
\hat{O}(t)=e^{\tilde{\mathcal{L}}_{\mathrm{mf}}^{\ddag}t}\hat{O}(0)\,,
\end{align*}
where $\tilde{\mathcal{L}}_{\mathrm{mf}}^{\ddag}$ is the adjoint of $\tilde{\mathcal{L}}_{\mathrm{mf}}$ that fulfills $
\mathrm{Tr}(\hat{O}\tilde{\mathcal{L}}_{\mathrm{mf}}\hat{\varrho})=\mathrm{Tr}(\tilde{\mathcal{L}}_{\mathrm{mf}}^{\ddag}\hat{O}\hat{\varrho})$ for operators $\hat{O}$ and $\hat{\varrho}$. 

Below we determine the matrix elements for the stationary state $\hat\rho_0$ (see Eq.~\eqref{inco}). For this specific state the matrix in Eq.~\eqref{CMatrix} becomes diagonal and the stability analysis is reduced to solving the equation
\begin{align}
1+C_{11}(s)=0\,.\label{1+c11}
\end{align}
In fact, for $X=0$ and in the reference frame rotating with $\omega_w$ we can write
\begin{align}
\hat{J}_1(t)=\hat{\sigma}{\rm e}^{{\rm i}{\rm e}^{{\rm i}\chi}\frac{w}{2}t}\cos(k\hat{x}+k\hat{p}t/m)\,.\label{J1}
\end{align}

We now calculate the value of $C_{11}$. Using Eq.~\eqref{C11} and Eq.~\eqref{J1} we get
\begin{align}
C_{11}=&{\rm i}\frac{\Lambda}{2}{\rm e}^{{\rm i}\chi}\int_{0}^{\infty} dte^{-st+{\rm i}{\rm e}^{{\rm i}\chi}\frac{w}{2}t}\left\langle\cos(k\hat{x})\cos\left(k\hat{x}+\frac{k\hat{p}}{m}t\right)\right\rangle\\
&={\rm i}\frac{\Lambda}{2}{\rm e}^{{\rm i}\chi}\int_{0}^{\infty} dte^{-st+i{\rm e}^{{\rm i}\chi}\frac{w}{2}t+{\rm i}\omega_Rt}I
\end{align}
where we have explicitly used that all particles are in the excited state (therefore $\langle\hat{\sigma}\hat{\sigma}^{\dag}\rangle=0$ holds). Moreover, we have introduced 
\begin{align}
I&=\int_{-\infty}^{\infty} dp\bigg\langle p\bigg|\cos(k\hat{x})\frac{e^{{\rm i}k\hat{x}}e^{{\rm i}k\hat{p}/mt}+e^{-{\rm i}k\hat{x}}e^{-{\rm i}k\hat{p}/mt}}{2} f(\hat{p})\bigg|p\bigg\rangle\nonumber\\
&=\frac{1}{4}\int_{-\infty}^{\infty} dp\langle p|f(\hat{p})|p\rangle (e^{{\rm i}kp/mt}+e^{-{\rm i}kp/mt})+\frac{1}{4}\left(\langle \hbar k\big|f(\hat{p})|-\hbar k\rangle +\langle -\hbar k\big|f(\hat{p})|\hbar k\rangle\right)e^{-{\rm i}2\omega_Rt}.\label{I}
\end{align}
where we have used that $
e^{{\rm i}k\hat{x}+{\rm i}k\hat{p}/mt}=e^{{\rm i}k\hat{x}}e^{{\rm i}k\hat{p}/mt}e^{{\rm i}\omega_Rt}$, with $\omega_R=\hbar k^2/2m$ the recoil frequency, and $e^{{\rm i}k\hat{x}}|p\rangle=|p+\hbar k\rangle$.
Using Eq.~\eqref{I} we obtain
\begin{align}
C_{11}=&i\frac{\Lambda\left(s-{\rm i}{\rm e}^{{\rm i}\chi}\frac{w}{2}-i\omega_R\right)}{4}{\rm e}^{{\rm i}\chi}\int_{-\infty}^{\infty} dp\frac{\langle p|f(\hat{p})|p\rangle}{\left(s-{\rm i}{\rm e}^{{\rm i}\chi}\frac{w}{2}-i\omega_R\right)^2+\left(\frac{kp}{m}\right)^2}\nonumber\\
&+{\rm i}\frac{\Lambda}{8}{\rm e}^{{\rm i}\chi}\frac{\langle \hbar k\big|f(\hat{p})|-\hbar k\rangle +\langle -\hbar k\big|f(\hat{p})|\hbar k\rangle}{s-{\rm i}{\rm e}^{{\rm i}\chi}\frac{w}{2}+{\rm i}\omega_R}.\label{C11finalhom}
\end{align}
For a thermal state (see Eq.~\eqref{homGaussian}) the dispersion relation then reads
\begin{align}
1+{\rm i}\frac{\Lambda y}{4\sin(\chi)}\mathrm{e}^{\mathrm{i}\chi}\int_{-\infty}^{\infty} dp\sqrt{\frac{\beta}{2m\pi}}\frac{\exp\left(-\beta\frac{p^2}{2m}\right)}{y^2+\left(\frac{kp}{m}\right)^2}=0\label{Homogeneousgaussiandispersion}
\end{align}
where $y=s-\mathrm{i}\mathrm{e}^{\mathrm{i}\chi}w/(2\sin\chi)-{\rm i}\omega_R$.  Using the substitution $p=\bar{p}u$ with $\bar{p}=\hbar k\Lambda/(2\omega_R)$ we can cast Eq.~\eqref{Homogeneousgaussiandispersion} into the form given by Eq.~\eqref{hogaussdis}.

\end{widetext}

\end{document}